\documentclass[sigconf,nonacm]{acmart}

\usepackage[normalem]{ulem}
\usepackage{booktabs}
\usepackage{colortbl}
\usepackage{xcolor}
\usepackage{array}
\usepackage{tabularx}
\usepackage{multirow} 
\usepackage{microtype}
\usepackage{graphicx}
\usepackage{subfigure}
\usepackage{booktabs} % for professional tables
\usepackage{hyperref}
\usepackage{extramarks}
\usepackage{enumitem}
\usepackage{fancyhdr}
\usepackage{booktabs}
\usepackage[table]{xcolor}
\usepackage{array}
\usepackage{listings}
\usepackage[dvipsnames]{xcolor}
\usepackage{float}
\usepackage{tabularx}
\usepackage{array}
\usepackage{xcolor}
\usepackage{tikz}

\newcommand{\name}{\textsc{AromaGen}}

\author{Awu Chen}
\affiliation{%
  \institution{MIT Media Lab}
  \city{Cambridge, MA}
  \country{USA}
}
\email{awwu@mit.edu}

\author{Silang Wang}
\affiliation{%
  \institution{University of Washington}
  \city{Seattle, WA}
  \country{USA}
}
\email{swang233@uw.edu}

\author{Yunge Wen}
\affiliation{%
  \institution{MIT Media Lab}
  \city{Cambridge, MA}
  \country{USA}
}
\email{yungew@mit.edu}

\author{Jianing Yu}
\affiliation{%
  \institution{Harvard Graduate School of Design}
  \city{Cambridge, MA}
  \country{USA}
}
\email{nomy_yu@gsd.harvard.edu}

\author{Zinab Eisa}
\affiliation{%
  \institution{Harvard GSD/SEAS}
  \city{Cambridge, MA}
  \country{USA}
}
\email{zeisa@mde.harvard.edu}

\author{Ana Julia Santos Lopes}
\affiliation{%
  \institution{Bowdoin College}
  \city{Brunswick, ME}
  \country{USA}
}
\email{alopes@bowdoin.edu}

\author{Jas Brooks}
\affiliation{%
  \institution{MIT CSAIL}
  \city{Cambridge, MA}
  \country{USA}
}
\email{jasb@mit.edu}

\author{Hiroshi Ishii}
\affiliation{%
  \institution{MIT Media Lab}
  \city{Cambridge, MA}
  \country{USA}
}
\email{ishii@mit.edu}

\author{Paul Pu Liang}
\affiliation{%
  \institution{MIT Media Lab}
  \city{Cambridge, MA}
  \country{USA}
}
\email{ppliang@mit.edu}

\begin{document}

\title{\name: Interactive Generation of Rich Olfactory Experiences with Multimodal Language Models}

\begin{teaserfigure}
  \centering
  \includegraphics[width=1\textwidth]{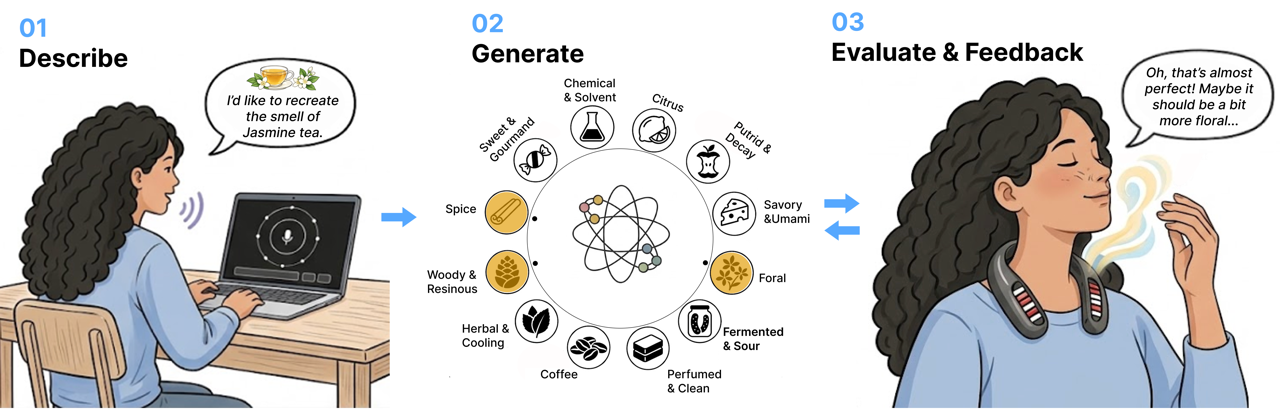}
    \caption{\name\ maps a free-form natural-language description of a desired olfactory experience to a 12-dimensional odorant ratio vector over 12 base odorants, physically rendered through a neck-worn olfactory display, and supports iterative refinement through free-form natural-language feedback.
  }
  % \vspace{2mm}
  \label{fig:teaser}
\end{teaserfigure}

\begin{abstract}
Smell’s connections with food, memory, and social experience have motivated researchers to bring olfaction into interactive systems. Multimodal AI opens new possibilities for generating smells from natural language, yet it remains unclear whether pretrained models encode sufficient olfactory knowledge to translate language into perceptible compositions. We present AromaGen, an AI-powered 12-channel wearable olfactory system that maps free-form natural-language descriptions to 12 base odorants selected to cover a semantically derived olfactory space, while supporting iterative refinement through natural-language feedback. In a between-subjects study (N=60) across 50 real-world benchmark smells, participants distinguished AromaGen-generated target smells above chance in a three-alternative forced-choice task, with retrieval-augmented AI composition performing comparably to human-expert composition and zero-shot AI composition. Natural-language feedback significantly improved the perceived similarity of both AI- and human-expert compositions. Our findings demonstrate the feasibility of using pretrained multimodal AI, with human aroma composition data, to generate perceptible olfactory experiences from natural language.
\end{abstract}

\maketitle



%\keywords{} shall we use keywords instead? a minor formatting issue
\iffalse
\Terminology
\name\ composition AromaGen composition (our approach best version) 
\name\ zero-shot composition AromaGen zero-shot composition
expert-derived composition (human-expert composition) 
describable olfactory space
12-base odorants 
12-dimensional odorant-ratio vector
human aroma composition data
Floral, Citrus, Woody /& Resinous, Herbal /& Cooling, Spice,
Sweet /& Gourmand, Roasted /& Smoky, Fermented /& Sour, Putrid /&
Decay, Chemical /& Solvent, Savoury /& Umami
\fi

\section{Introduction}

Smell is deeply entangled with the human experience. It shapes how we taste food~\cite{bultInvestigationsMultimodalSensory2007}, recall memories~\cite{herzOdorMemoryReview1996}, and even form social bonds~\cite{fruminSocialChemosignalingFunction2015}. These qualities have long motivated Human-Computer Interaction (HCI) researchers and artists to bring smell into interaction design, from immersive VR~\cite{brooksTrigeminalbasedTemperatureIllusions2020,ranasingheSeasonTravellerMultisensory2018} and gaming systems~\cite{ranasingheTaintedOlfactionenhancedGame2019,choiScentedDaysExploring2024} to health~\cite{riera2017sense} and wellbeing~\cite{amoresDevelopmentStudyEzzence2022}. Despite decades of work, olfactory interfaces remain constrained to fixed scent cartridges or reservoirs \cite{yamanakaStudyRecordingApple2003,yamanakaStudyOdorRecorder2003, hopper2024multi,dobbelstein2017inscent} and pre-defined generation patterns \cite{prasetyawanOdorReproductionTechnology2024}, while general-purpose reproduction of a broad range of real-world smells from a compact odorant basis remains an open challenge \cite{prasetyawanOdorReproductionTechnology2024}.

\textit{AI for smell} broadly refers to the use of AI models to classify or generate olfactory information by mapping between different representations of smell, such as language, sensor measurements, molecular structures, and odorant compositions. On the classification side, AI has been used to recognize odors from physical measurements, including electronic-nose sensor arrays~\cite{feng2025smellnet} and chemical measurements such as GC--MS \cite{8489539}, as well as to predict olfactory properties directly from molecular structure~\cite{leePrincipalOdorMap2023}. On the generation side, AI could translate a desired olfactory experience into a composition that can be physically rendered~\cite{10943150, nakamoto2026overview, clarke_anemoia_2025}. Chemoinformatics and optimization-based approaches typically start from molecular or sensor-derived targets~\cite{10943150, nakamoto2026overview}, whereas recent tangible systems map language-mediated semantics to blends over a fragrance library~\cite{clarke_anemoia_2025}. Specifically, the application of generative AI to olfaction could potentially make olfactory interfaces substantially more expressive \cite{nakamoto2026overview}, allowing users to transform natural language-based description of olfactory experiences into physically realizable scent mixture, rather than requiring designers to manually specify aromas in advance. Emerging research suggests that multimodal LLMs encode cross-modal relationships between linguistic descriptions and olfactory perception~\cite{kurfali2025}, raising the possibility of using their pretrained knowledge to directly translate natural-language smell descriptions into physically realizable compositions, without retraining a dedicated olfactory generative model. However, two fundamental questions remain unresolved. First, unlike color or sound, smell has no established set of ``primaries'' analogous to RGB from which arbitrary percepts can be composed ~\cite{gottfried2010central}, leaving unclear what primary olfactory basis an LLM could use to generate different smells. Second, although existing work suggests that LLMs encode meaningful olfactory information, it remains unclear whether this knowledge is sufficient to bridge semantic smell descriptions and physical compositions that humans perceive as corresponding to the intended smells. These gaps motivate our investigation of whether a multimodal LLM, paired with a compact olfactory basis, can support general-purpose aroma generation directly from natural language.

Building on prior tangible systems that map language-mediated semantics to blends over a fixed fragrance library~\cite{clarke_anemoia_2025}, we present AromaGen, a general-purpose aroma-generation system based on generative AI that translates everyday descriptions of smells into compositions over a compact olfactory basis of 12 base odorants, and we ask whether such compositions are sufficiently perceptible and target-associated to support general-purpose use at scale. In addition to utilizing the latent olfactory knowledge of a pre-trained multimodal model, AromaGen specializes the model for smell generation by providing olfactory-specific context, human-generated aroma composition data, optional iterative natural language-based feedback, and retrieval-augmented generation (RAG) over previously evaluated aroma formulations. This addresses a central challenge in AI-based aroma generation: although foundation models encode relationships between language and olfactory perception, they have not been trained to translate arbitrary smell descriptions into formulations for a particular physical odorant palette. Because olfaction has no agreed set of primaries~\cite{gottfried2010central}, AromaGen's 12 channels are instead aligned to 12 olfactory clusters derived from a semantic-space PCA of English smell descriptors (UMBC olfactory subcorpus; pipeline adapted from Hörberg~\cite{horberg2022}), providing the compositional space within which the model operates. A neck-worn dispenser physically renders the resulting mixtures, while users could iteratively evaluate and refine them through natural language-based feedback (e.g., `make it stronger'' or `less sweet''), allowing subsequent compositions to incorporate human perceptual judgments. Together, these components turn a general-purpose multimodal LLM into an olfactory composition system grounded in smell-specific knowledge, human odor composition data, and a physically realizable odorant space. 

We evaluated AromaGen using 50 real-world benchmark smells spanning the 12 olfactory clusters. Our primary measure is a three-alternative forced-choice (3-AFC) task, which tests whether a generated smell is sufficiently associated with its intended real-world target for participants to distinguish it from two unrelated distractors. Here, by \textit{association}, we mean that the generated smell carries sufficient olfactory information for participants to relate it to the intended target over other unrelated smells, for example, recognizing a generated smell containing citrus and vanilla tones as corresponding more closely to an orange than to blue cheese or garlic. It therefore measures target recognition rather than one-to-one perceptual indistinguishability. In a between-subjects study, we compared AI-generated compositions with formulations independently constructed by olfactory experts, alongside direct ratings of perceived similarity of the AI-composed smell to the physical target. Participants in the interactive conditions could then provide optional natural-language feedback and re-rate the resulting smell, allowing us to evaluate whether interactive refinement improves perceived similarity.

In summary, our contributions are as follows:
\begin{itemize}
\item An extension of the language-to-blend aroma composition paradigm introduced by \citet{clarke_anemoia_2025}: we specialize a multimodal LLM for physical smell generation over a structured 12-odorant output space by combining olfactory-specific context, human-generated odor composition data, retrieval-augmented generation over previously evaluated formulations, and optional iterative natural-language feedback. 
\item An AI-powered 12-channel wearable olfactory system whose base odorants are aligned to a PCA-informed covering of describable olfactory space, enabling AI-generated formulations of smell to be physically rendered and iteratively refined through natural-language feedback.
\item An evaluation across 50 real-world smell targets and 12 olfactory clusters that measures whether generated smells are recognizably associated with their intended targets, compares AI-generated and human expert compositions, and evaluates changes in perceived similarity following natural-language refinement.
\end{itemize}

Together, AromaGen demonstrates an approach for grounding the olfactory knowledge of multimodal LLM in smell-specific examples, human perceptual feedback, and a physically realizable odorant space, enabling users to move from open-ended descriptions of desired smells to interactively generated physical approximations.

\section{Related Work}

Our work builds upon research in multimodal AI, developing machine learning models for olfactory data, systems for aroma reproduction and approximation, and olfactory interfaces.

\subsection{Olfaction as an AI Modality}

The evolution of large language models from text-only systems to multimodal architectures has substantially expanded the range of modalities that AI can perceive and generate~\citep{liang2024foundations,tsai2019multimodal,wu2023multimodal}. Today's state-of-the-art multimodal models are capable of jointly processing text, images, video, audio, tactile, physical sensing, physiological data, and other interaction modalities~\citep{feng2025smellnet,kaur2025single,liang2025multimodal,song2025opentouch}. Despite trained only on language, even language models themselves have shown surprising commonsense knowledge about other modalities and domains, like robotics, social interactions, and physical embodied AI~\citep{liu2025aligning}. This has opened up many opportunities for leveraging foundation models in tangible, embodied, and multimodal interaction systems~\citep{duan2022survey}. Furthermore, recent advances in interactive AI agents~\cite{durante2024agent,qu2026coral,zhou2026mem1} provide an opportunity to design systems that continuously perceive and generate smells, learn from user feedback, and improve over time. Smell, however, remains largely absent from the AI landscape.

Olfactory perception begins when airborne molecules bind to receptors in the nose, producing signals the brain interprets as odor. To study this systematically, researchers have assembled datasets that characterize molecules along three layers: chemoinformatic structural representations, physicochemical properties (molecular weight, boiling point, vapor pressure, logP), and human perceptual ratings including semantic descriptors (e.g., \textit{floral}, \textit{fruity}), similarity judgments, and detection thresholds. The Dravnieks Atlas~\citep{dravnieks1982odor} established a foundational 146-descriptor space for monomolecular odorants analyzed with linear methods;~\citet{kellerOlfactoryPerceptionChemically2016} extended this with 476 molecules. More recently,~\citet{leePrincipalOdorMap2023} represented molecules as graphs and trained a message-passing neural network to derive a ``Principal Odor Map''. Work on aroma mixtures has remained largely descriptive~\citep{snitzPredictingOdorPerceptual2013, raviaMeasureSmellEnables2020}, and the scarcity of labeled perceptual data has historically constrained model complexity~\citep{gerkin2021parsing}. Large-scale sensor datasets such as SmellNet~\citep{feng2025smellnet} open new possibilities for foundation-model approaches to smell sensing.

Our work takes a different approach: rather than predicting perception from molecular features, we leverage the olfactory knowledge already encoded in multimodal LLMs to translate free-form language into actionable aroma formulations.

\subsection{Olfactory Language and Semantics}
A longstanding assumption in cognitive science holds that smell is uniquely resistant to verbal description, a phenomenon termed \textit{olfactory ineffability}~\cite{majidBurenhult2014}. Studies with English speakers confirm that odor talk is infrequent, dedicated smell terms are rare, and naming familiar aromas is surprisingly difficult even for people with intact olfaction~\cite{majid2021}. However,~\citet{majidBurenhult2014} demonstrated that this is not a universal human limitation: speakers of Jahai, a language spoken by hunter-gatherers on the Malay Peninsula, name aromas as easily as colors, suggesting that the smell-language gap is shaped by culture and ecology rather than biology~\cite{majidBurenhult2014}.

Separately,~\citet{horberg2022} took a corpus-based approach to map the structure of olfactory language in English, automatically identifying aroma descriptors from natural text and deriving their semantic organization~\cite{horberg2022}. Despite the sparse aroma vocabulary of English, they found a coherent semantic space structured primarily along dimensions of pleasantness and edibility, suggesting that latent olfactory knowledge is nonetheless encoded in natural language. This view is further supported by recent work showing that large language models, particularly GPT-4o, can recover olfactory-semantic relationships from text to a degree that correlates with human perceptual judgments~\cite{kurfali2025}. Complementary interactive work examines the same language--smell link under physical stimulation: in Sniff AI, \citet{zhong2024sniff} study ``sniff-and-describe'' tasks in which participants smell presented scents and an LLM attempts to infer the odor from their descriptions, reporting limited and uneven perceptual alignment across scents. We adapt Hörberg's descriptor-discovery pipeline to the UMBC WebBase olfactory subcorpus and run a distance-matrix PCA plus clustering to obtain a 12-cluster covering that the hardware palette and the 50-descriptor evaluation both follow. 

Together, these findings motivate our approach: rather than requiring sensor measurements or molecular specifications, we leverage the olfactory knowledge already latent in multimodal LLMs to bridge natural language and smell, and we pick the 12 channels from that language space rather than from three foods.

\subsection{Aroma Reproduction and Approximation}

Aroma reproduction research investigates how to recreate a target aroma by blending known components so that the resulting mixture is perceptually similar to the original. Unlike vision or audition, olfaction has no set of primaries from which arbitrary percepts can be synthesized~\citep{gottfried2010central}; reproduction therefore relies on searching a mixture space over a finite component palette.

\citet{nakamotoOlfactoryDisplayOdor2016} introduced the concept of an \textit{odor recorder} that uses QCM sensor arrays to capture a fruit odor's signature and then searches mixture ratios over a small component set to approximate it~\citep{munoz-aguirreOdorApproximationFruit2007, yamanakaStudyRecordingApple2003, nakamotoOlfactoryDisplayOdor2016}. For example, orange (originally produced from 14 ingredients) could be convincingly approximated with only three (linalool, citral, and decanal)~\citep{yamanakaStudyOdorRecorder2003}. Complex essential oils have similarly been approximated using 10--30 components with non-negative matrix factorization and mass spectrometry~\citep{nakamotoOdorApproximationUsing2012, prasetyawanOdorReproductionTechnology2024}. More recent algorithmic frameworks combine odorant descriptor prediction with optimization:~\citet{nicolaiRealTimeAroma2016} demonstrated real-time aroma reconstruction of citrus varieties using 12 components, and deep learning approaches combine mass spectra with descriptor prediction and gradient descent to design new formulations~\citep{aleixandreAutomaticScentCreation2024}.

A critical limitation shared by all these approaches is that they require either sensor measurements of a physical target sample or precise molecular specifications as input: none accept the open-ended natural-language descriptions that people use in everyday life to communicate about smell.

\subsection{Olfactory Interfaces}

Work on olfaction in HCI has primarily focused on \emph{output}: designing devices that release aromas to enrich interactive experiences. Motivated by smell's strong links to memory, emotion, and presence~\citep{hartmannTripJapanSixteen1902, heiligSensoramaSimulator1962}, researchers have explored applications spanning healthcare, entertainment, and social interaction---including breathing guidance~\citep{zhouEscentCoachWearable2025}, motion sickness reduction~\citep{schartmullerSickScentsInvestigating2020, reichlInvestigatingImpactOdors2024}, narrative enrichment~\citep{choiScentedDaysExploring2024}, social collaboration~\citep{mehrotraScentCollaborationExploring2022, liMidAirGesturesProactive2025}, and body perception alteration~\citep{brianzaQuintEssenceProbeStudy2022}.
Delivery mechanisms range from thermal diffusion and aerosolization~\citep{ranasingheSeasonTravellerMultisensory2018, yanagidaUnencumberingLocalizedOlfactory2003, amoresEssenceOlfactoryInterfaces2017} to directed airflow, ultrasound-steered vortices~\citep{hasegawaMidairUltrasoundFragrance2018}, and body-worn devices including on-face wearables~\citep{wangOnFaceOlfactoryInterfaces2020}, trigeminal nose clips~\citep{brooksStereoSmellElectricalTrigeminal2021}, and retronasal aroma displays for eating and drinking~\citep{liAromaBiteAugmentingFlavor2025, mayumiBubblEatDesigningBubbleBased2025}.

Despite this breadth, most olfactory interfaces still rely on a fixed repertoire of pre-selected aromas, precluding dynamic, user-specified aroma generation at runtime. Following the semantic-to-blend composition approach demonstrated by \citet{clarke_anemoia_2025}, \name\ couples an LLM-based formulation pipeline with a 12-channel olfactory display, enabling users to request a described aroma in natural language and receive a synthesized approximation on demand; we further specialize this pipeline with a PCA-aligned odorant basis, human aroma composition data, retrieval-augmented generation, and iterative natural-language refinement (Section~\ref{sec:interaction}).

\section{\name\ }

\name\ is an interactive, general-purpose olfactory generation system that maps natural-language descriptions of desired olfactory experiences onto compositions of 12 base odorants. While the system architecture can accommodate multimodal inputs, including text, speech, and images, in this work we focus specifically on natural-language input through text and speech. Text descriptions are entered directly through the \name\ interface, while spoken descriptions are transcribed verbatim before being passed to the same AI composition pipeline. We do not evaluate image-based aroma generation in this work.

Given an input description, \name\ infers a composition represented as a ratio vector over its 12 base odorants and renders the resulting mixture through its olfactory display. Users may optionally provide natural-language feedback describing how they would like the generated smell to change. \name\ can then use this feedback together with the preceding composition to generate a revised smell, allowing users to iteratively refine the olfactory experience when desired.

\subsection{Base Odorant Selection}

A central design decision in \name\ is the selection of the 12 base odorants that define its generative space, with the number of odorants constrained by our 12-channel olfactory display. Because users specify desired olfactory experiences in \name\ through natural language rather than chemical properties, we characterize the space of smells that \name\ aims to generate semantically rather than chemically. Specifically, we define the \textit{describable olfactory space} as the range of olfactory experiences that can be meaningfully expressed and differentiated through natural-language descriptions in English, and use this semantic space to guide the selection of a compact and diverse set of 12 base odorants.

\subsubsection{Constructing the Describable Olfactory Space}

To construct the describable olfactory space, we adapted the corpus-based approach of Horberg et al.~\cite{horberg2022}, which identifies odor-related vocabulary from natural English text and models its organization using distributional semantics. Using the 2.97-billion-token UMBC WebBase Corpus, we extracted 92,120 sentences associated with olfactory and gustatory seed terms (e.g., \emph{odor}, \emph{smell}, \emph{scent}, \emph{fragrance}, and \emph{aroma}) and identified 3,558 candidate olfactory descriptors based on their association with these contexts. We then learned semantic representations of these descriptors using Word2Vec and applied PCA to characterize the dominant dimensions of the resulting olfactory semantic space; full details of descriptor extraction, model training, and robustness analyses are provided in Appendix~\ref{app:olfactory-space}.

\begin{figure*}
    \centering
    \includegraphics[width=\linewidth]{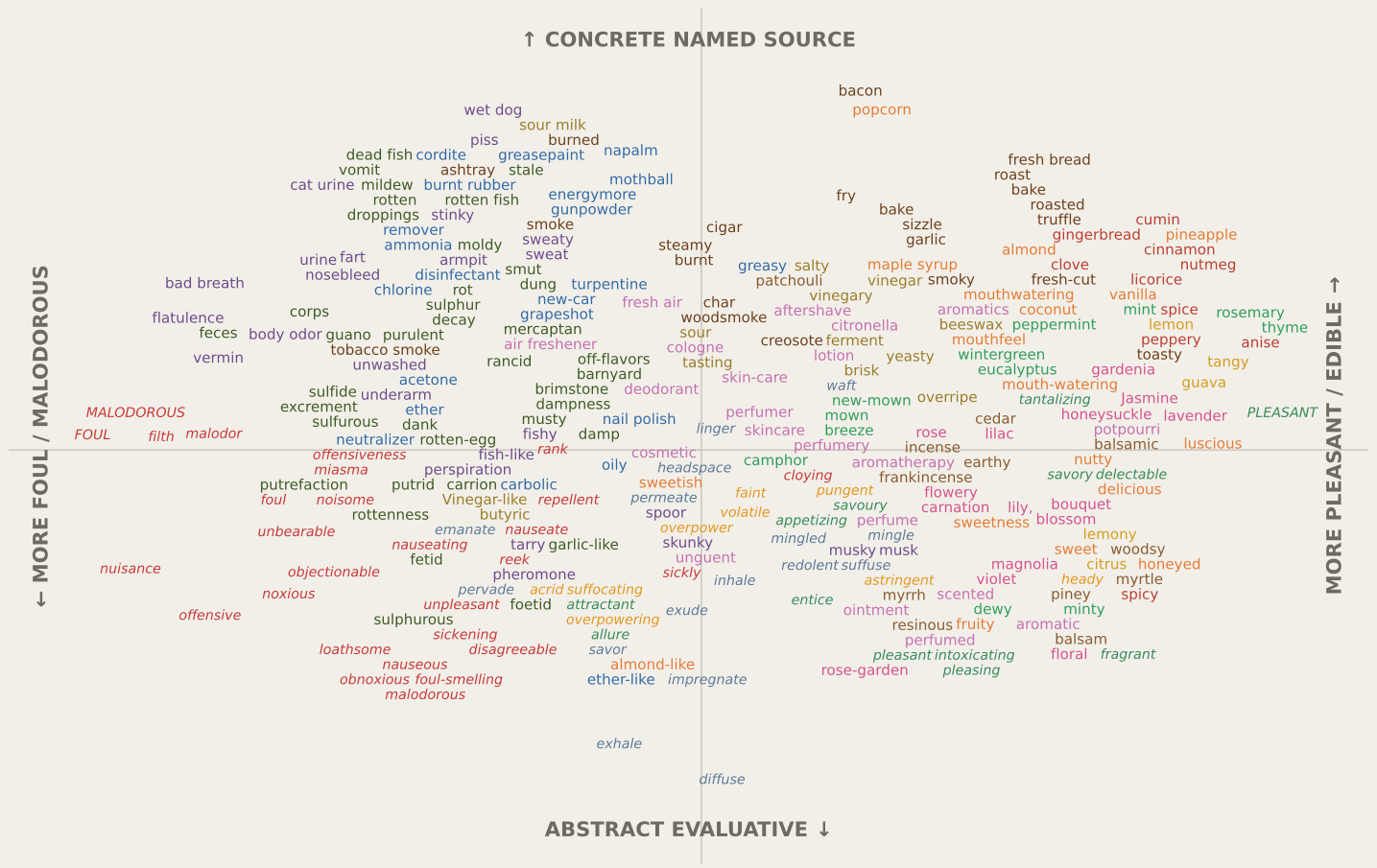}
        \caption{Two-dimensional describable olfactory space of English smell descriptors, grouped into 12 perceptually informed clusters (same color indicates the same cluster). X-axis represents the pleasantness / unpleasantness axis, Y-axis represents the concrete odor source / abstract evaluative axis.}
    \label{fig:olfactory space}
\end{figure*}

Across different thresholds for selecting olfactory-associated vocabulary upon which the semantic-based PCA was carried out, two stable and perceptually-meaningful semantic dimensions emerged. The first primarily differentiated pleasant from unpleasant olfactory descriptions, while the second differentiated concrete odor sources from more abstract or evaluative descriptions. We retained these two dimensions as the two principal axis of \name's describable olfactory space, as shown in Figure~\ref{fig:olfactory space}. 

\subsubsection{From Semantic Space to Initial Base Odorants}

We used the 317 descriptors within the top 10\% OAI threshold \cite{horberg2022} of our candidate olfactory descriptors to populate the constructed describable olfactory space as shown in Figure~\ref{fig:olfactory space}. Two researchers independently reviewed these descriptors and manually grouped them according to similarities in the olfactory experiences they described. The researchers subsequently compared their groupings, discussed disagreements, and consolidated them into 12 perceptually-informed clusters spread across the four quadrants of the describable olfactory space: \emph{Floral}, \emph{Citrus}, \emph{Woody \& Resinous}, \emph{Herbal \& Cooling}, \emph{Spice}, \emph{Sweet \& Gourmand}, \emph{Roasted \& Smoky}, \emph{Fermented \& Sour}, \emph{Putrid \& Decay}, \emph{Chemical \& Solvent}, \emph{Body \& Animalic}, and \emph{Perfumed \& Clean}.

We then consulted two professional perfumers to select an initial representative odorant for each cluster. At this stage, each of \name's 12 base odorants corresponded to a single essential oil intended to capture the characteristic olfactory quality of its corresponding cluster.

\subsubsection{Refining the 12 Base Odorants}

We subsequently conducted internal testing to examine whether the initial basis provided sufficient coverage within each cluster. From the 317 descriptors retained at the 10\% OAI threshold, we selected 50 descriptors with direct physical referents that could be reproduced as reference odor sources. This initial benchmark spanned all 12 clusters, with four descriptors from ten clusters and five descriptors from two clusters. Examples included \emph{jasmine}, \emph{tobacco}, and \emph{fresh bread}. We used these physical references to qualitatively assess \name's ability to approximate different smells within each cluster using the initial 12 base odorants.

This testing revealed that for some clusters, a single representative odorant could sometimes capture the general character of a cluster while remaining too specific to represent its broader range of descriptors. For example, the initial \emph{Roasted \& Smoky} cluster was solely represented by the birch tar oil essential oil. While it provided a strong smoky and charred character suitable for some targets, its distinctive barbecue-like quality made it less suitable for other descriptors within the same cluster, such as \emph{coffee} and \emph{fresh bread}. Based on these observations, we conducted a second round of consultation with a third olfactory expert in perfumery, who evaluated the initial 12 base odorants through \name\ and recommended combining complementary essential oils for clusters where a single material provided limited coverage. We therefore converted six of the initial 12 base odorants into fixed composites of two or three constituent essential oils to increase their olfactory representativeness.

Internal testing also informed another change to the 12 base odorants. The initial taxonomy contained a \emph{Body \& Animalic} cluster, which included descriptors such as \emph{vomit}, \emph{urine}, \emph{droppings}, \emph{feces}, and \emph{excrement}. Because our subsequent evaluation required participants to directly smell real-world references of the smells under evaluation alongside their \name-generated approximations, exposing participants to real-world representation of such smells in the laboratory raised practical, hygiene, and participant-safety concerns. We therefore excluded this cluster from the 12 base odorants and subsequent user studies. At the same time, internal testing suggested that savory and umami qualities were insufficiently represented by the existing 12 base odorants when approximating several real-world targets. A chef participating in our internal testing independently identified this limitation and recommended introducing a dedicated savory and umami base odorant that could remind people of smells produced from the Mailard reaction \cite{elhosry2025maillard}. Due to reasons stated above, we eventually replaced the \emph{Body \& Animalic} cluster with a \emph{Savoury \& Umami} cluster in our finalized 12 base odorants.

Following these refinements, we finalized \name's 12 base odorants, as shown in Table~\ref{tab:palette}. Each hardware channel corresponded to one of the 12 final perceptually-informed clusters and contained either a single representative essential oil or a fixed composite of two or three complementary essential oils. The exact ways how \name\ produce synthetic smells from these channels with essential oils are detailed in Section~\ref{sec:interaction}. This iterative process allowed us to broaden the range of olfactory qualities represented by the 12 base odorants and ensure that the finalized set of base odorants could be evaluated using real-world odor references, all within the constraint of our 12-channel hardware limitations. 

\begin{table*}[t]
\centering
\normalsize
\setlength{\tabcolsep}{4pt}
\renewcommand{\arraystretch}{1.08}

\begin{tabularx}{\linewidth}{@{}
    c
    >{\raggedright\arraybackslash}X
    c
    >{\raggedright\arraybackslash}p{0.40\linewidth}
    >{\raggedright\arraybackslash}p{2.85cm}
@{}}
\toprule
\textbf{\#} & \textbf{Odorant} & \textbf{Cluster} & \textbf{Notes} \\
\midrule
1  & Benz Sal & Perfumed / Clean & Soft floral, powdery, balsamic \\
2  & Sandalwood & Woody / Resinous & Creamy, warm wood \\
3  & Clove Bud (50\%) + Cumin (50\%) & Spice & Warm pungent clove, earthy cumin \\
4  & Lavender (50\%) + Rose (50\%) & Floral & Herbaceous lavender, dewy rose \\
5  & Orange (50\%) + Lemon (50\%) & Citrus & Sweet-tart citrus \\
6  & Vanilla Sugar (50\%) + Almond Extract (50\%) & Sweet / Gourmand & Creamy vanilla, marzipan almond \\
7  & Birch Tar (33.3\%) + Coffee (33.3\%) + Clove Bud (33.3\%) & Roasted / Smoky & Tarry smoke, roasted coffee, clove \\
8  & Eucalyptus & Herbal / Cooling & Camphoraceous, medicinal-fresh \\
9  & Cognac & Chemical / Solvent & Boozy, solvent-like pungency \\
10 & Vinegar & Fermented / Sour & Sharp acetic sourness \\
11 & Isovaleric Acid & Animal / Body & Sweaty, cheesy, animalic \\
12 & Seaweed (40\%) + Fenugreek (40\%) + Garlic (20\%) & Savory / Umami & Marine umami, curry-like fenugreek, garlic \\
\bottomrule
\end{tabularx}

\caption{The 12 base odorants in the live \name\ palette. The palette was selected to span diverse regions of the describable olfactory space under the 12-channel hardware constraint. Composite odorants were prepared by combining their constituent materials in the proportions shown.}
\label{tab:palette}
\Description{A 12-row palette table. Each row has a channel number, the odorant or composite blend, the matching olfactory cluster, and a short sensory note. Six rows are single materials and six are fixed blends of two or three oils.}
\vspace{-2mm}
\end{table*}

\subsection{AI Modeling of Aroma Composition}

Given \name's 12 base odorants, we formulated aroma composition as a structured prediction task in which an LLM maps a natural-language description of a desired olfactory experience onto a mixture of the available base odorants, following the semantic-to-olfactory translation pipeline established in prior tangible work~\cite{clarke_anemoia_2025}. \name\ adds three complementary mechanisms on top of that core mapping: AromaGen zero-shot composition from a user's description, retrieval-augmented composition informed by previously evaluated generations, and iterative refinement through natural-language feedback.

Formally, given a natural-language input $x$, the model predicts a ratio vector
\[
\mathbf{r}=(r_1,\ldots,r_d), \qquad r_i \geq 0,\qquad \sum_{i=1}^{d}r_i=1,
\]

where $d=12$ and each $r_i$ represents the relative contribution of one base odorant. The resulting ratio vector is subsequently converted by the hardware controller into an interleaved odor-delivery sequence, as described in Section~\ref{sec:interaction}.

\subsubsection{AromaGen Zero-shot Generation}

Our baseline approach generates compositions zero-shot, without examples of previously evaluated smells or their odorant ratios. The model receives a fixed system prompt specifying the 12 base odorants and the qualitative olfactory characteristics of each base odorant, together with composition constraints and the required structured output format. The prompt instructs the model to translate the user's requested olfactory experience into ratios over the available bases, typically using between two and six active odorants. The same base system prompt is used for both AromaGen zero-shot and retrieval-augmented composition; the latter differs only in the addition of retrieved, previously evaluated examples as contextual evidence. The complete system prompt is provided in Appendix~\ref{app:prompt}.

For both typed and spoken interactions, the composition model ultimately receives a natural-language description: text is passed directly to the model, whereas speech is transcribed verbatim before inference. Although \name's architecture can accommodate additional input modalities, we evaluate only text- and speech-based natural-language input in this work.

\begin{figure}
    \centering
    \includegraphics[width=\linewidth]{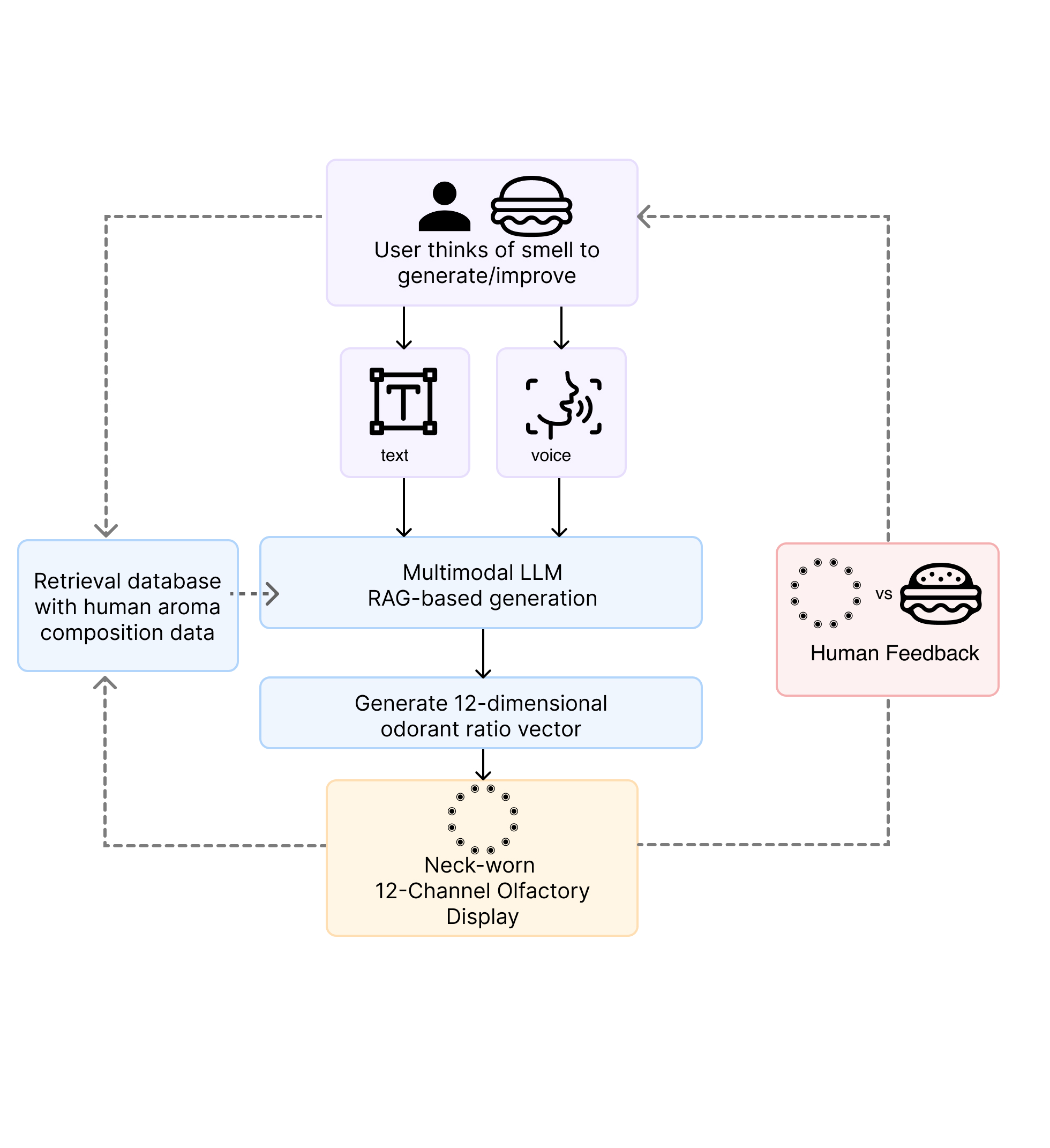}
    % \vspace{-2mm}
    \caption{The \name\ composition pipeline (extending semantic-to-blend translation~\cite{clarke_anemoia_2025}). Users describe a desired olfactory experience through text or speech. The model combines pretrained knowledge with retrieved human aroma composition data to produce a 12-dimensional odorant-ratio vector over the 12 base odorants. The wearable renders the mixture, after which users can iteratively refine it through natural-language feedback.}
    \label{fig:system-pipeline}
    % \vspace{-2mm}
\end{figure}

\subsubsection{Retrieval-Augmented Composition}

Zero-shot generation with system prompt provides the LLM with semantic knowledge of the available base odorants, but no empirical information about how compositions produced by \name's specific 12 base odorants are perceived by humans. To provide this perceptual grounding, we additionally constructed a retrieval database of human aroma composition data from human evaluations of a 50 real-world smell benchmark.

We constructed a second benchmark of 50 real-world smells (Table~\ref{tab:benchmark}) for the retrieval database and subsequent evaluation. The benchmark was organized around the 12 perceptually-informed olfactory clusters established during base odorant selection, with four target smells selected from ten clusters and five from two clusters. Within each cluster, we selected target smells to span different levels of olfactory complexity, including relatively simple smells dominated by a single recognizable olfactory quality (e.g., orange and garlic) and more compositionally complex smells comprising multiple olfactory qualities (e.g., cinnamon roll and Korean barbecue beef patty). All targets were selected to have physical real-world referents that could be practically prepared and presented in the lab during subsequent evaluation and comparison with \name's generated smells.

To construct the database of human aroma composition data of this 50 real-world smell benchmark, three researchers independently evaluated \name-generated compositions for all the smells in this benchmark. For the operationalization and preparation of the 50 benchmark smells, please refer to Section~\ref{sec:benchmark-smell-preparation} For each benchmark smell, the researcher first smelled the physical reference odor source and then smelled the corresponding composition generated by \name. They rated the perceptual similarity between the physical and generated smells on a 7-point Likert scale and could provide natural-language feedback describing how the generated smell should be changed. \name then generated a revised composition based on this feedback, which the researcher smelled and evaluated again. This process could be repeated until the researcher was satisfied with the generated smell or did not feel that further feedback would improve it. Across all three researchers, this process produced a collection of human aroma compositions data spanning the 50-smell benchmark.

Each entry in the resulting retrieval database associates a target smell with its 12-dimensional odorant-ratio vector, its human-rated perceptual similarity to the physical target on a 7-point scale, and any corresponding natural-language feedback. Unlike general linguistic or olfactory knowledge available during pretraining, these data provide empirical evidence about how particular mixtures produced by \name's 12 base odorants were perceived by human evaluators.

With the RAG component, whenever \name\ is given a new target description, it retrieves the top-5 semantically similar evaluated examples from this database and provides them to the LLM as contextual evidence. Retrieval does not simply reproduce the composition of the nearest example: retrieved targets may differ from the current request, and the LLM uses their formulations, ratings, and feedback as precedents for constructing a new ratio vector. Highly rated compositions (6--7 on similarity scale) are presented as strong positive precedents, compositions rated 5 on similarity scale as moderate precedents that may be adapted, and lower-rated compositions (1--4 on similarity scale) primarily as negative evidence whose associated feedback can indicate what should be changed or avoided. In this way, RAG supplements the LLM's pretrained semantic knowledge with human perceptual data specific to \name's physical generation space, without requiring model retraining.

The retrieval database can further grow through interaction. After a generated composition is physically rendered and rated, its target description, base odorant ratio vector, similarity rating, and optional natural-language feedback are added as a new retrieval entry. Subsequent generations can therefore draw on an ever expanding history of human-evaluated aroma compositions as \name\ is used to generate more and more smells.

\begin{table}[t]
\centering
\small
\begin{tabularx}{\linewidth}{p{2.5cm} X}
\toprule
\textbf{Olfactory Cluster} & \textbf{Real-World Smell Targets} \\
\midrule

Floral &
Lavender; rose; jasmine tea; peony and rose oil shampoo \\

Citrus &
Orange; mango; lemonade; lime soda (Sprite) \\

Woody \& Resinous &
Birch; patchouli; whiskey and oak candle; incense \\

Herbal \& Cooling &
Basil; cucumber; peppermint tea; mint chewing gum \\

Spice &
Ginger; black pepper; chai tea; cinnamon roll \\

Sweet \& Gourmand &
Coke; dark chocolate; apple pie; sweet popcorn; chocolate and marshmallow Pop-Tarts \\

Roasted \& Smoky &
Coffee beans; bacon; Korean barbecue beef patty; hot dog with hot sauce \\

Fermented \& Sour &
Greek yogurt; pickled cucumber; fries with ranch sauce; nachos with sour cream \\

Putrid \& Decay &
Blue cheese; durian; canned sardines; natto beans \\

Chemical \& Solvent &
Whiskey; tequila; mint fluoride mouthwash; lavender nail polish remover \\

Perfumed \& Clean &
Aloe vera; hand sanitizer; mint fluoride toothpaste; almond oil shampoo \\

Savoury \& Umami &
Soy sauce; Parmesan cheese; garlic; seasoned pulled pork in barbecue sauce; salty popcorn \\

\bottomrule
\end{tabularx}
\caption{The 50 real-world smell targets used in the comparative user study, organized by \name's 12 olfactory clusters.}
\vspace{-2mm}
\label{tab:benchmark}
\end{table}

\subsubsection{Iterative Refinement}

While RAG incorporates human aroma composition data accumulated from previous generations, retrieved precedents may still not fully capture the current user's intended smell or individual perception. \name\ therefore closes the perceptual gap by allowing users to evaluate the physical output and refine it through natural-language feedback. After smelling a generated composition, a user can describe how they would like it to change, for example, requesting that it be less smoky or more citrus-like, or accept the current generation without further refinement.

At refinement iteration $t$, the LLM receives the original target description, the current ratio vector $\mathbf{r}^{(t-1)}$, the preceding interaction history, and the user's latest feedback. Rather than generating a new composition independently, the model is instructed to revise the existing formulation: preserving aspects that were not criticized, adjusting relative weights implicated by the feedback, and introducing or removing odorants when appropriate. The resulting ratio vector $\mathbf{r}^{(t)}$ is rendered by \name\ and can be smelled, evaluated, and refined again, creating an iterative \textit{describe--generate--smell--refine} loop.

\begin{figure}
    \centering
    \includegraphics[width=\linewidth]{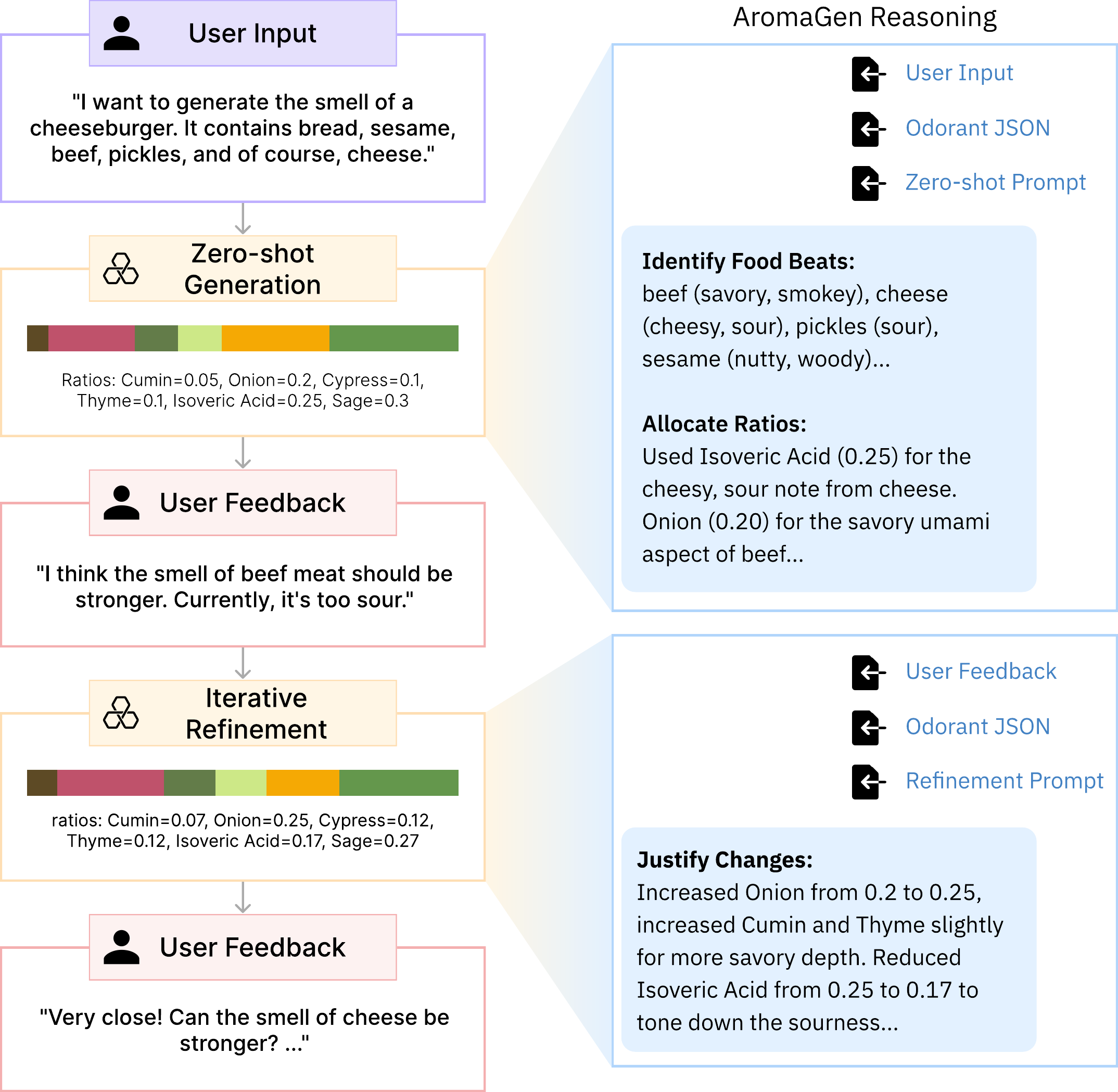}
    \caption{An example of \name's iterative refinement: the model decomposes a natural-language request, maps it onto perceptual qualities, and allocates a 12-dimensional odorant-ratio vector over the 12 base odorants. Spoken feedback (e.g., ``less sour'') is applied as an in-context revision of the current mixture.
    }
    \vspace{-2mm}
    \label{fig:example}
\end{figure}

Throughout this process, \name\ logs the target description, generated ratio vectors, similarity evaluations, natural-language feedback, and iteration history. These interactions provide both immediate context for within-session refinement and new human aroma composition data that can be incorporated into the retrieval database for future generations. Together, RAG and iterative refinement allow \name\ to complement the LLM's latent olfactory knowledge with accumulated human evaluations of aroma composition specific to \name's 12 base odorants, while adapting individual compositions through direct perceptual feedback. Figure~\ref{fig:example} shows an example of this describe--generate--smell--refine loop, including how natural-language feedback updates the 12-dimensional odorant-ratio vector.

\begin{figure}
    \centering
    \includegraphics[width=\linewidth]{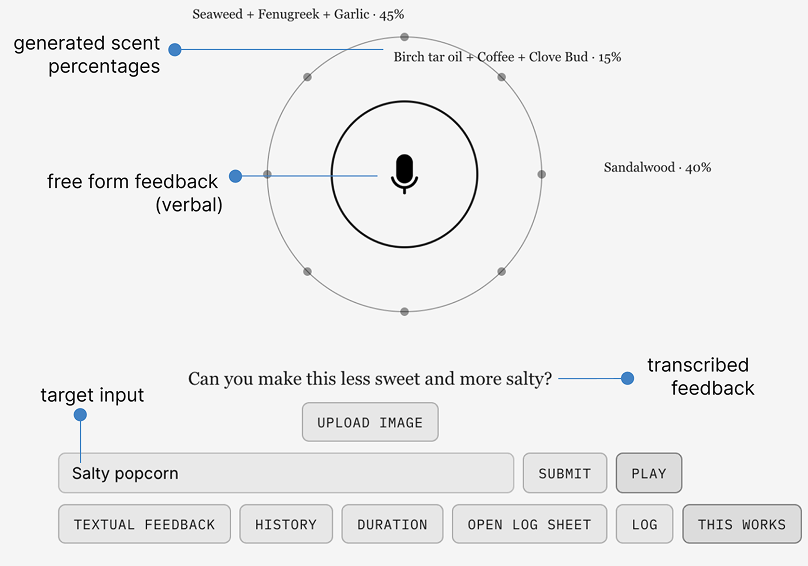}
    \vspace{-3mm}
        \caption{Experimenter-facing web interface for \name: natural-language input (text or speech), a visualization of the 12-dimensional odorant-ratio vector, playback on the wearable, and natural-language feedback. Image input is supported by the architecture but was not evaluated.}
    % \vspace{-4mm}
    \label{fig:interface}
\end{figure}

\begin{figure}
    \centering
    \includegraphics[width=\linewidth]{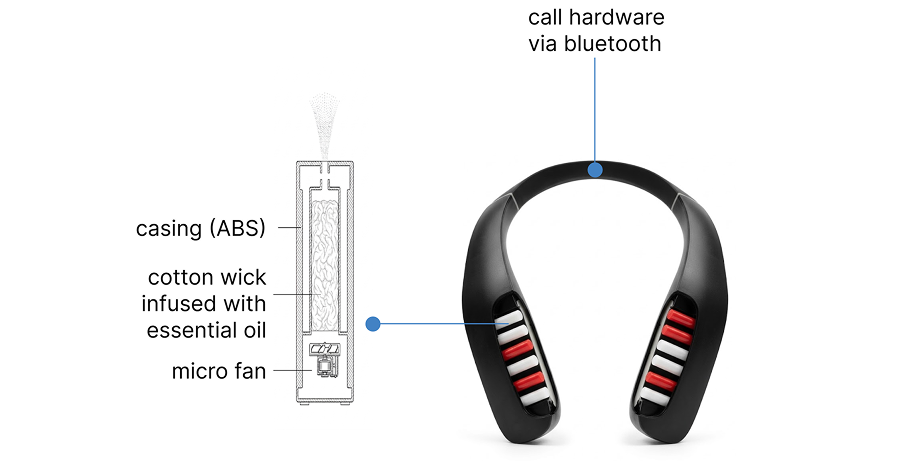}
    \caption{\name\ hardware, with 12 cartridges containing 12 base odorants}
    \label{fig:hardware}
\end{figure}

\subsection{Interaction Design}
\label{sec:interaction}

The interaction design of \name\ follows three goals: enabling users to specify desired olfactory experiences through natural language rather than manually manipulating odorant ratios, translating AI-generated compositions into physical smells through a compact multichannel display, and supporting iterative refinement based on users' perceptual feedback. Together, these components form a closed-loop interaction in which a user can \textit{describe, generate, smell, and refine} an olfactory experience. We describe below the web interface through which this interaction is mediated, the wearable olfactory display used to physically render compositions, and the pulse-based delivery mechanism that translates continuous odorant ratios into device actuation.

\subsubsection{Interface Design}

\name\ UI is implemented as a web-based interface (Figure~\ref{fig:interface}) consisting of an input panel and a circular visualization of the 12-odorant composition. Users can describe a desired smell through typed text or speech. Spoken input is automatically transcribed and passed verbatim to the same natural-language composition pipeline used for typed input. While the broader system architecture supports image input, image-based aroma generation was not evaluated in this work. In the scope of this paper, we only allow experimenters to use the \name\ UI, and any input from the participants are either typed into \name\ UI or transcribed verbatim into the same system.

After generation, the interface visualizes the selected base odorants and their predicted relative weights. Users can dispatch the composition to the olfactory display for presentation. After smelling the result, they may optionally enter or speak natural-language feedback describing how they would like the smell to change, triggering the iterative refinement pipeline described above.

\subsubsection{Hardware Implementation}

\name\ is built upon the SCENTAC X-Scent 3.0, a neck-worn digital scent display housing 12 independently controlled channels (DC 5,V, $<380$,g), as shown in Figure~\ref{fig:hardware}. We replaced the original 12 channels with custom cartridges containing essential oil-soaked perfume test strips, corresponding to each of \name's 12 base odorants. Each channel uses a fan to direct airflow through the cartridges and can be independently actuated through channel-specific BLE commands. The predicted composition is transmitted from the \name\ interface to the hardware controller as a 12-dimensional ratio vector, which the controller converts into an interleaved sequence of odor pulses as described below.

\subsubsection{Odor Delivery Mechanism}

As the device is limited in its ability to mix scents due to the hardware constraint that only one channel could be activated at a time, \name\ renders an AI-predicted composition as an interleaved, ratio-weighted sequence of short odor pulses. Let $\mathbf{r}=(r_1,\ldots,r_{12})$ denote the predicted composition and $A={i \mid r_i>0}$ the set of active odorants, with $m=|A|$. Given $R$ delivery rounds, the controller allocates a total pulse budget
\begin{equation}
    P = Rm,
\end{equation}
where $R=10$ in our current implementation. This design scales the number of pulses with the number of constituent odorants in a composition. The controller allocates these $P$ pulses across the active odorants approximately in proportion to their predicted ratios. Thus, for odorant $i$, its allocated number of pulses $p_i$ approximates
\begin{equation}
    p_i \approx r_iP,
\end{equation}
subject to integer allocation across the available pulse budget. Each pulse actuates the corresponding odorant channel for $\delta=0.5$s. Consequently, odorants assigned larger ratios receive proportionally more pulses rather than a single longer period of continuous actuation.

The allocated pulses are distributed approximately evenly throughout the delivery sequence, interleaving contributions from the constituent odorants rather than dispensing each odorant as a single contiguous block. Because the total pulse budget is proportional to the number of active odorants, the duration of a presentation varies with composition complexity rather than following a fixed dispensing cycle. A composition containing more distinct odorants therefore requires a longer presentation than one containing fewer odorants. This pulse-based mechanism provides a discrete physical approximation of the continuous ratio vector predicted by the AI while distributing each constituent throughout the generated olfactory experience.

\section{Methods}

Having established \name's 12 base odorants, AI-driven aroma composition pipeline and interaction design, we next evaluated the system through user studies. The primary goal of our studies was to examine whether \name\ could perceptually recreate all the smells in the 50 real-world smell benchmark with its 12 base odorants, and how its AI-generated compositions compared with those constructed by human experts. We further investigated how natural-language feedback affected generated smells and explored how experienced users engaged with \name\ when unconstrained by predefined target odors.

Overall, our studies aimed to addresses the following research questions:

\begin{itemize}

\item \textbf{RQ1 --- Synthetic olfactory representation:} To what extent are \name-composed smells perceived as similar to their corresponding targets in the 50 real-world smell benchmark?

\item \textbf{RQ2 --- AI composition:} How do \name-composed smells compare with expert-composed smells in 3-AFC identification accuracy and perceived similarity?

\item \textbf{RQ3 --- Interactive refinement:} To what extent can natural-language feedback improve the perceived similarity of generated smells to their intended targets?

\item \textbf{RQ4 --- Free-form generation:} How do users perceive and explore \name's capabilities when generating smells beyond the 50 real-world smell benchmark?

\end{itemize}

Together, these studies evaluate \name\ at three levels: the representational capacity of \name's fixed 12 base odorants, the performance of its AI composition and refinement mechanisms relative to human-expert compositions, and the ways experienced users explore the system when given open-ended control.

We explored each of the research questions via three studies. First, we conducted a structured expert elicitation study \cite{ohagan2006uncertain} ($N=3$) in which three olfactory experts independently composed all the smells in the 50 real-world smell benchmark using \name's fixed 12 base odorants. This established a human expert baseline under the same physical constraints as \name's AI composer. Next, we conducted our primary comparative user study ($N=60$), comparing \name\ compositions, human expert compositions, and \name\ zero-shot compositions using a three-alternative forced-choice (3-AFC) olfactory discrimination test ~\cite{10943150}, in which participants compared each of the 12 target smells in the study with three other sequentially presented smells to identify the one that best matched the target, after which they performed similarity ratings between the real and synthetic target smell, and optional feedback on the synthetic smell if they are in the RAG and expert conditions. Finally, we conducted a small-scale exploratory free-form aroma generation study ($N=5$), examining the system's perceived capabilities and limitations, perceived affordances, and opportunities for creative olfactory exploration beyond the 50 real-world smell benchmark. The study protocol was approved by the Institutional Review Board (IRB; approval number: 2507001742), and we obtained all participants' consent before their participation. 

\subsection{Expert Elicitation Study}
\label{sec:expert-study}

To contextualize \name's AI-generated aroma compositions against human expertise, we asked how olfactory experts would reconstruct the same 50 real-world smells when constrained to \name's 12 base odorants. We therefore conducted a structured expert elicitation study in which three olfactory experts independently composed all 50 benchmark smells using the same fixed 12 base odorants available to \name. Rather than treating expert formulations as perceptual ground truth, this study established a human expert baseline for comparing AI and Human-expert composition under identical physical constraints.
\subsubsection{Participants}

We recruited three experts (two women and one man) with professional experience involving the composition and perception of complex aromas. Expert 1 was an AI researcher at a U.S.-based university studying human taste, with prior professional experience as a barista composing coffee beverages with distinct flavor and aroma profiles. Expert 2 was an executive pastry chef overseeing the pastry departments of multiple restaurants, with professional experience composing flavors and aromas for baked goods using ingredients such as coffee extract and vanilla sugar. Expert 3 was a perfumer and store manager at a U.S. perfumery specializing in custom scent creation, where she regularly composed personalized fragrances using scent materials including citrus and sandalwood essential oils. Together, the experts provided complementary perspectives spanning sensory research, culinary practice, and perfumery.

\begin{figure}[t]
    \centering
    \includegraphics[width=\linewidth]{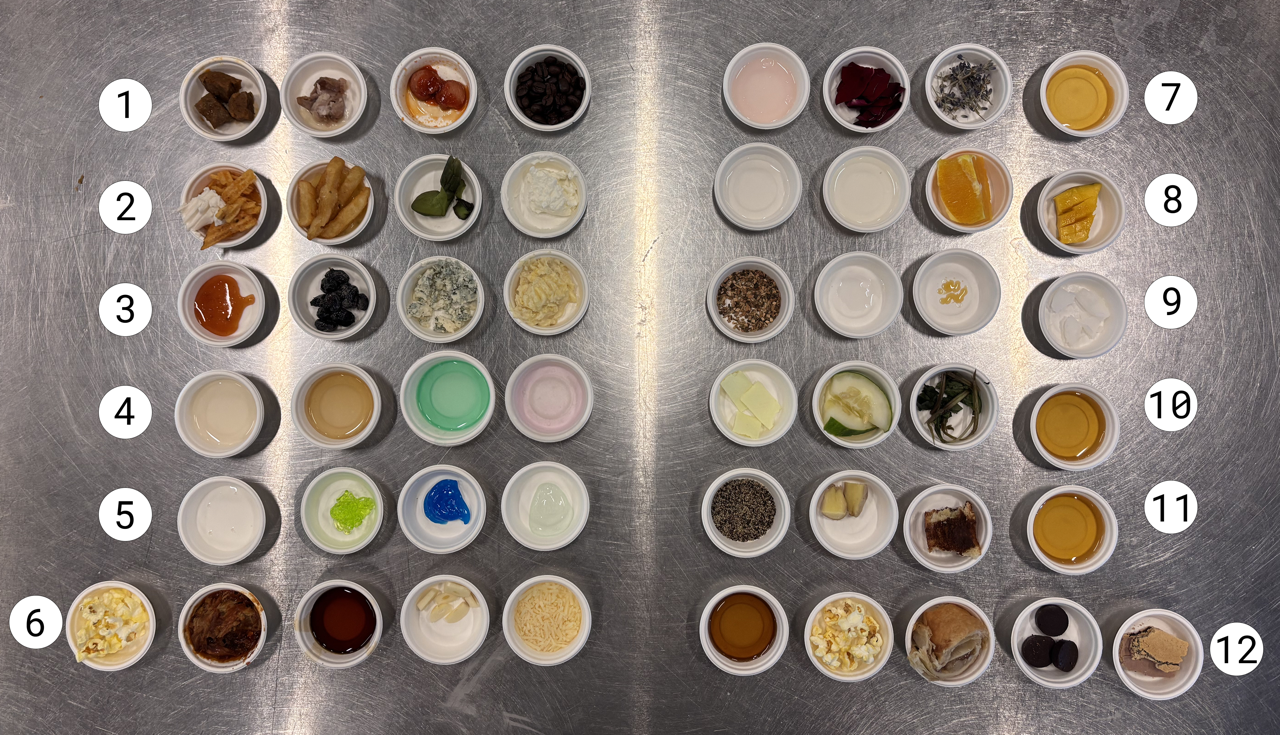}
    
    \caption{The 50 physical odor sources used as references for our real-world benchmark smells across 12 olfactory clusters, presented in odorless paper portion cups. Each row corresponds to one olfactory cluster, with odor sources ordered from left to right.}
    \label{fig:physical-odor-sources}
    
    \vspace{2mm}
    \footnotesize
    \begin{tabularx}{\linewidth}{@{}X@{\hspace{4mm}}X@{}}
        \textbf{Left (top to bottom, left to right)} & \textbf{Right (top to bottom, left to right)} \\[2pt]
        
        \textbf{1. Roasted \& Smoky:} Korean barbecue beef patty, bacon, 
        hot dog with hot sauce, coffee beans
        &
        \textbf{7. Floral:} peony and rose shampoo, rose petals, 
        lavender flower, jasmine tea
        \\[3pt]
        
        \textbf{2. Fermented \& Sour:} nachos with sour cream, fries with ranch sauce, 
        pickled cucumber, Greek yogurt
        &
        \textbf{8. Citrus:} lime soda, lemonade, navel orange, mango
        \\[3pt]
        
        \textbf{3. Putrid \& Decay:} canned sardines, natto beans, blue cheese, durian
        &
        \textbf{9. Woody \& Resinous:} incense, birch essential oil, 
        patchouli essential oil, whiskey and oak candle
        \\[3pt]
        
        \textbf{4. Chemical \& Solvent:} whiskey, tequila, mint fluoride mouthwash, 
        lavender nail polish remover
        &
        \textbf{10. Herbal \& Cooling:} mint chewing gum, cucumber, basil leaves, 
        peppermint tea
        \\[3pt]
        
        \textbf{5. Perfumed \& Clean:} almond oil shampoo, aloe vera, 
        mint fluoride toothpaste, hand sanitizer
        &
        \textbf{11. Spice:} black pepper, ginger, cinnamon roll, chai tea
        \\[3pt]
        
        \textbf{6. Savoury \& Umami:} salty popcorn, seasoned pulled pork in barbecue sauce, 
        soy sauce, garlic, Parmesan cheese
        &
        \textbf{12. Sweet \& Gourmand:} Coke, sweet popcorn, apple pie, dark chocolate, 
        chocolate marshmallow Pop-Tart
    \end{tabularx}
\end{figure}

\subsubsection{Benchmark Smell Preparation}
\label{sec:benchmark-smell-preparation}
We operationalized the 50 benchmark smells in the scope of this paper with 50 physical odor sources corresponding to each of the benchmark smell, as shown in Figure~\ref{fig:physical-odor-sources}. To maintain the freshness and perceptibility of the 50 physical odor sources throughout the study, the experimenter prepared them one hour before the first study participant each day. All physical odor sources were presented in odorless squat paper portion cups and were refreshed at least daily, with more volatile or perishable odor sources prepared more frequently according to their material properties.

Liquid odor sources, such as jasmine tea, lime soda, and tequila, were freshly poured into a new cup each time they were used in a study session. Edibles containing meat or oil, such as hot dogs with hot sauce, french fries with ranch sauce, and Korean barbecue beef patties, were reheated in a microwave immediately before presentation. More stable solid odor sources, such as cucumber, garlic, and blue cheese, were freshly prepared at the beginning of each study day and replaced as needed.

Immediately before presenting each physical odor source to a participant, two experimenters independently smelled the physical odor sources to verify that its characteristic odor was readily detectable. If either experimenter judged the odor to be insufficiently detectable, the odor source was replaced with a freshly prepared sample before presentation. This procedure was used throughout all relevant studies or internal testing to reduce variation in physical reference odors resulting from evaporation, cooling, or prolonged exposure to the environment.

\subsubsection{Base Odorant Preparation and Expert Onboarding}

We prepared the 12 base odorants individually in separate 4mL capped glass vials, each labeled with its corresponding olfactory cluster. Bases represented by a single odorant contained only that odorant, whereas multi-odorant bases were prepared by mixing their constituents in the specificed in Table~\ref{tab:palette}, using a Toppette single-channel pipettor. For example, the ``Savoury / Umami'' base contained 40\% of seaweed, 40\% fenugreek, and 20\% garlic essential oils by volume. We also replaced the essential oil-soaked perfume test strips in each of the 12 cartridges of \name\ with a freshly prepared one to ensure \name\ could produce each base odorant clearly.

Before elicitation, experts smelled each base odorant first from its reference vial and then individually through the corresponding \name\ cartridge to familiarize themselves with the system. All three experts described the smells from the vial presentation as more intense but otherwise reported no noticeable differences in their perceived olfactory quality.

\subsubsection{Elicitation Procedure}

Each expert independently composed all 50 benchmark smells, whose presentation order was randomized for each expert. The 12 reference vials remained available throughout the session, allowing experts to conveniently inspect individual base odorants as needed while constructing their compositions.

For each benchmark smell, we first presented its physical odor source, before asking experts to specify a combination of base odorants and their relative ratios (e.g., ``30\% floral and 70\% putrid'') to approximate that target, with ratios summing to 100\%. They could use any number of the 12 base odorants in any combination, although ratios were typically specified in increments of five percentage points. These ratios controlled the relative actuation durations of \name's cartridges, as specified in Section~\ref{sec:interaction}.

Once the expert came up with a composition, the researcher entered each proposed composition into \name, and the expert compared the generated smell with the physical target and rated their similarity on a 7-point Likert scale. Experts could iteratively adjust the relative ratios, remove base odorants, or introduce additional base odorants. This generate--compare--rate--refine process continued until the expert felt further refinement would not lead to meaningful improvement in the generated smell. The experts were encouraged to converge within three refinement rounds to limit session duration and repeated odor exposure.

Once an expert was satisfied with an approximation, we recorded its 12-dimensional odorant-ratio vector as their final composition for that target, before moving on to the next target smell until all 50 benchmark smells have been composed.
To mitigate olfactory fatigue, experts took a 10-minute break after every 50 minutes of aroma composition, during which they left the laboratory and walked in an open-air environment. A handheld electronic fan was also available throughout the session to disperse residual odors between evaluations.

\subsubsection{Expert Elicitation Results}
Each of the three experts independently composed all 50 benchmark smells, yielding 150 candidate formulations. On average, the experts took 3.33h to complete the study, with the fastest being 2h, and the slowest being 4.5h. Across the three experts, while their raw 7-point similarity ratings showed some agreement on categories and targets that could be composed successfully from the 12 base odorants, their lowest-rated compositions were more heterogeneous. (Table~\ref{tab:expert-ratings}).

\begin{table*}[t]
\centering
\small
\begin{tabularx}{\linewidth}{@{}
    c
    >{\raggedright\arraybackslash}p{2.4cm}
    c
    >{\raggedright\arraybackslash}X
    >{\raggedright\arraybackslash}p{2.8cm}
@{}}
\toprule
\textbf{Expert} & \textbf{Background} & \textbf{Overall} & \textbf{Highest-rated clusters} & \textbf{Lowest-rated cluster} \\
\midrule
1 & Barista and taste researcher & 4.22 (1.36) & Chemical \& Solvent 5.75; Citrus 5.50 & Fermented \& Sour 2.50 \\
2 & Pastry chef & 5.40 (0.90) & Chemical \& Solvent 6.50; Citrus 5.75; Woody \& Resinous 5.75 & Roasted \& Smoky 4.50 \\
3 & Perfumer & 3.86 (0.93) & Citrus 4.50; Woody \& Resinous 4.50 & Putrid \& Decay 2.75 \\
\bottomrule
\end{tabularx}
\caption{Mean 7-point similarity of each expert's final compositions. Overall is the mean (SD) across all 50 targets. Cluster values are that expert's mean rating for targets in the named cluster.}
\label{tab:expert-ratings}
\Description{A three-row table of expert elicitation ratings. Expert 1, a barista and taste researcher, averaged 4.22. Expert 2, a pastry chef, averaged 5.40. Expert 3, a perfumer, averaged 3.86. The highest-rated clusters were Chemical and Solvent, Citrus, and Woody and Resinous. The lowest-rated clusters were Fermented and Sour, Roasted and Smoky, and Putrid and Decay.}
\end{table*}

Greater variation emerged among the lower-rated clusters. Expert 1 rated \textit{Fermented \& Sour} lowest ($M=2.50$), Expert 2 rated \textit{Roasted \& Smoky} lowest ($M=4.50$), and Expert 3 rated \textit{Putrid \& Decay} lowest ($M=2.75$). At the individual-target level, coffee beans received Expert 1's lowest rating (1/7), durian and salty popcorn received Expert 2's lowest ratings (3/7), and durian, canned sardines, apple pie, and dark chocolate received Expert 3's lowest ratings (2/7).

Because experts differed in their use of the 7-point similarity scale (Expert 1: \(M=4.22, SD=1.36\); Expert 2: \(M=5.40, SD=0.90\); Expert 3: \(M=3.86, SD=0.93\)), we also normalized similarity ratings within each expert using z-scores. For each of the 50 benchmark smells, we then selected the formulation with the highest z-normalized similarity rating among the three experts as the human-expert formulation for the subsequent comparative user study.

The final expert baseline of the 50 benchmark smells drew nearly equally from all three experts: 17 formulations (34\%) were contributed by Expert 1, 16 (32\%) by Expert 2, and 17 (34\%) by Expert 3. Thus, no single expert dominated the resulting baseline. We treat these selected formulations as a human expert baseline representing what trained practitioners could compose under \name's fixed 12-odorant constraint, rather than as perceptual ground truth for the target smells. During the Human-expert composition + Feedback condition in our comparative user study, all 50 smells from our benchmark were deterministically produced from \name\ following this expert baseline.

\subsection{Comparative User Study}

\begin{figure*}[t]
    \centering
    \includegraphics[width=\linewidth]{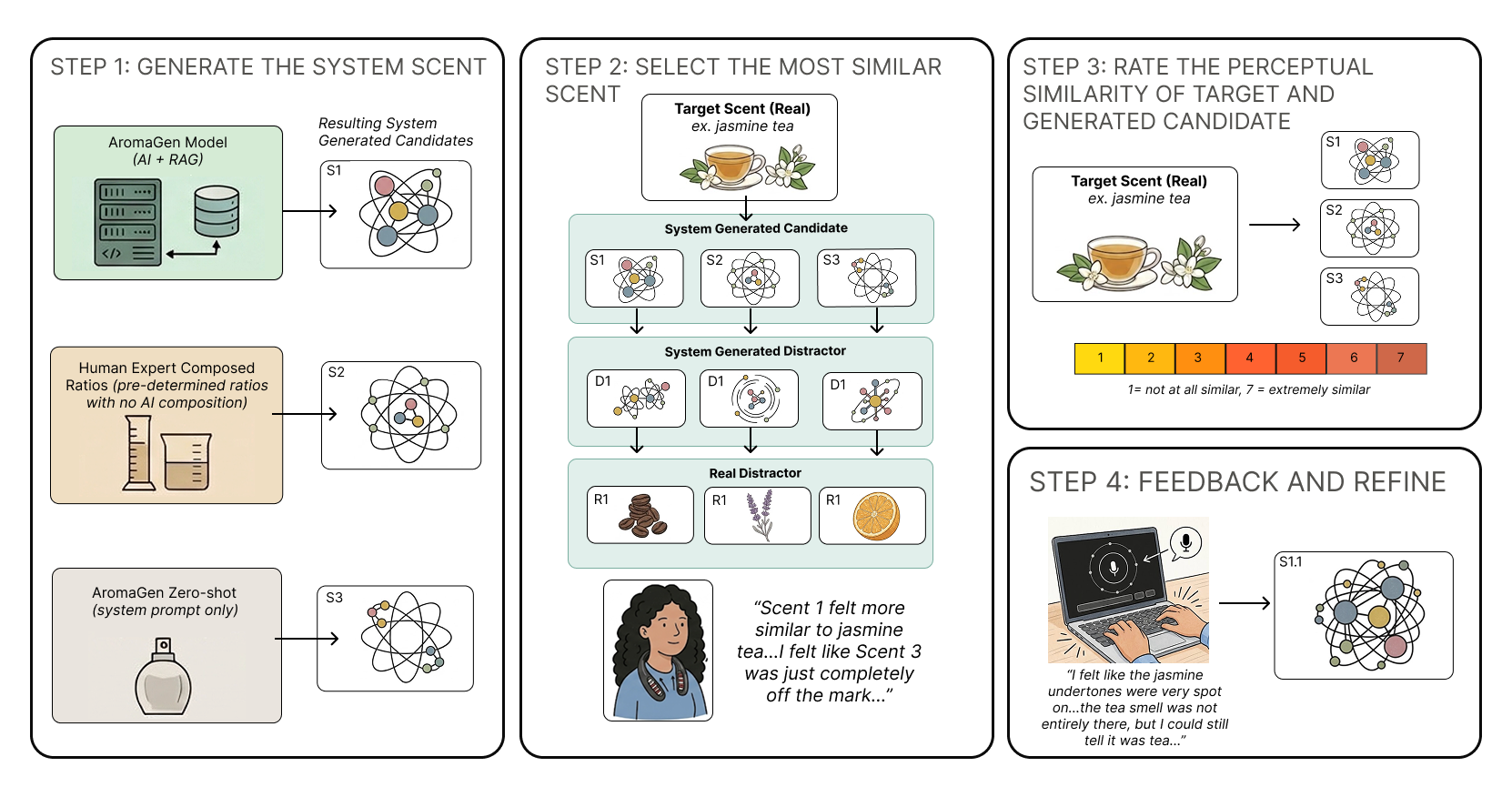}
    \caption{Overview of the comparative user study procedure across \name\ composition, human-expert composition, and \name\ zero-shot composition.}
    \label{fig:study-overview}
\end{figure*}

Our expert elicitation study established what trained olfactory practitioners could compose when constrained to \name's 12 base odorants. We next asked whether an LLM operating under the same physical constraints could produce smells that were perceptually comparable to this human expert baseline, and what retrieval augmentation and natural language feedback contribute to this process.

We addressed these questions through a between-subjects comparative user study of three smell composition approaches: \name\ composition, human-expert composition, and \name\ zero-shot composition. Rather than requiring generated smells to be perceptually indistinguishable from their real-world counterparts, we evaluated whether they preserved sufficient olfactory information for users to associate it with their intended targets and how similar participants perceived them to those targets. We therefore used two complementary measures: a three-alternative forced-choice (3-AFC) olfactory discrimination task and a direct 7-point rating of perceived similarity between each generated smell and its physical target. Our 3-AFC task was adapted from the 2-AFC evaluation of AI-generated scents by Aleixandre et al.~\cite{10943150}. Based on internal pilot testing, we selected three alternatives as a balance between the ease of distinguishing two alternatives and the increased memory demands of comparing four. After the 3-AFC test and similarity ratings, participants could optionally engage in natural language feedback to improve the synthetic smells. A visual summary of our comparative user study flow can be found in Figure~\ref{fig:study-overview}.

\subsubsection{Participants}

We recruited 62 participants from two U.S.\ university communities, of whom 60 completed the study and were included in the final analysis, with 20 participants in each condition. Two participants did not complete the study: Participant 4 was unable to complete the study within the one-hour session, while Participant 44 reported discomfort shortly after beginning the study and chose to discontinue participation. Their incomplete data were excluded from analysis.

Participants received \$25 for a one-hour, in-person study session. During participant recruitment, we excluded individuals who reported anosmia or other olfactory impairments, asthma, sensitivity to essential oils or perfumes, or allergies to any odor source or base odorant used in the study. Each participant took part in exactly one condition. Data collection proceeded sequentially by condition: \name\ composition, followed by human-expert composition, and then \name\ + zero-shot composition. Participant demographics are reported in Appendix~\ref{app:participant-demographics}.

\subsubsection{Study Design}

We used a between-subjects design comparing three approaches to composing target smells using the same \name\ hardware and 12 base odorants:

\begin{itemize}
    \item \textbf{\name\ + Feedback ($n=20$):} \name\ generated the initial composition using our RAG-based LLM pipeline. The retrieval database contained previous human aroma composition data and was incrementally updated with feedback collected during the study. Participants could optionally refine the generated composition through natural-language feedback.

    \item \textbf{Human-expert Composition + Feedback ($n=20$):} The initial composition was the fixed human-expert formulation established through our expert elicitation study (Section~\ref{sec:expert-study}). Participants could subsequently refine this composition through the same natural-language feedback mechanism used in the \name\ + feedback condition.

    \item \textbf{\name\ zero-shot ($n=20$):} \name\ generated the initial composition using the same LLM, system prompt, temperature, and output representation as the \name\ + feedback condition, but without retrieved examples. This condition served as an ablation of retrieval and did not include natural-language refinement.
\end{itemize}

Each participant completed 12 trials, evaluating one target from each of \name's 12 olfactory clusters. Targets were sampled from the 50-smell benchmark (Table~\ref{tab:benchmark}), with each benchmark smell represented by a corresponding physical reference odor prepared and maintained using the same procedure described in Section~\ref{sec:benchmark-smell-preparation}. Targets, distractors, and trial order were balanced across participants; full sampling procedures are provided in Appendix~\ref{app:comparative-sampling}.

\begin{figure*}
    \centering
    \includegraphics[width=\linewidth]{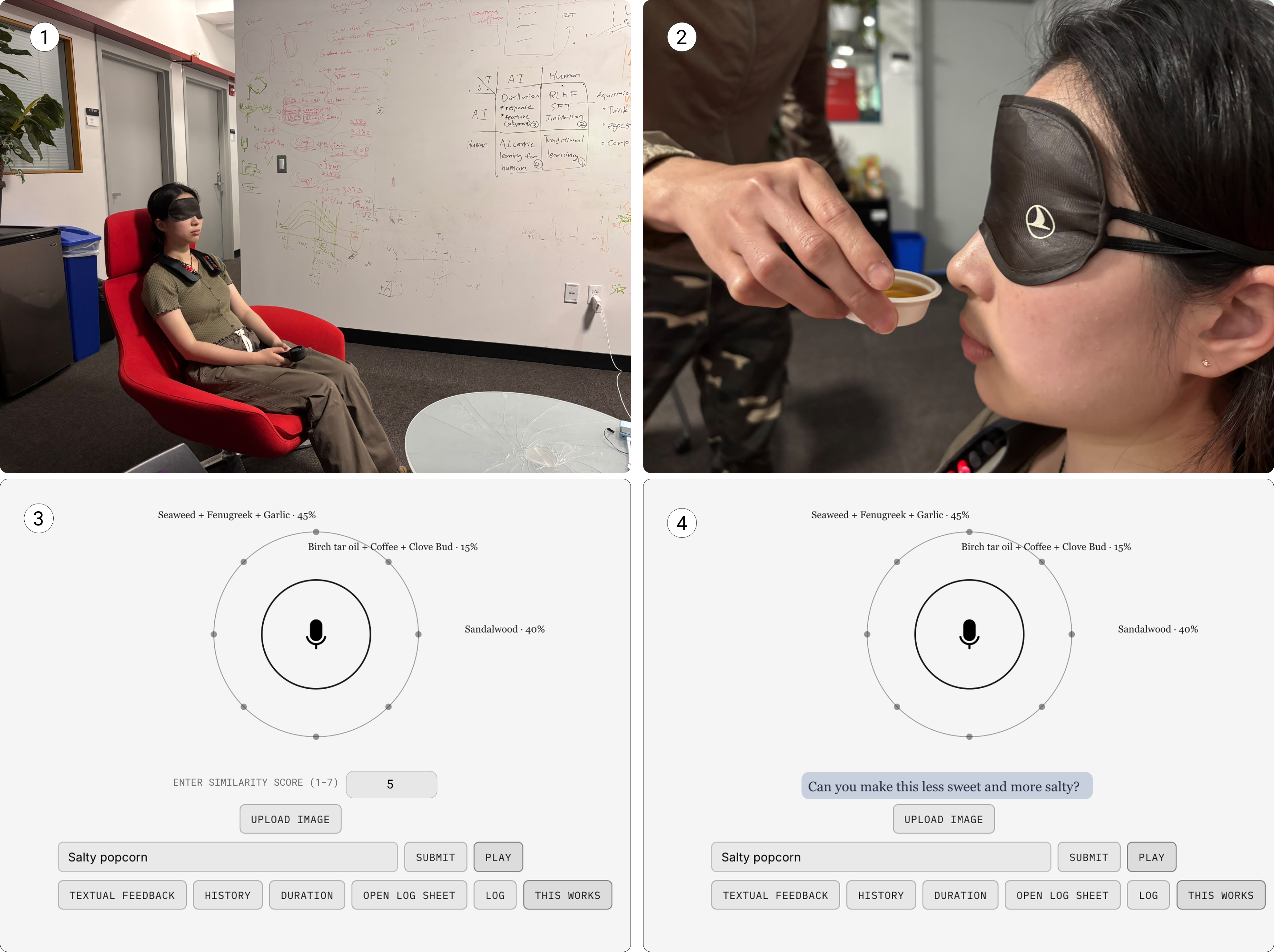}
    \caption{Comparative user study procedure. 1. Study environment. 2. Physical odor presentation. 3. Logging similarity ratings in \name\ interface. 4. Capturing natural language based feedback in \name\ interface.}
    \label{fig:user-study-procedure}
\end{figure*}

\subsubsection{Procedure}

Participants remained in a well-ventilated area and were blindfolded throughout the study. Physical odor sources were manually presented near the participant's nose in odorless squat paper portion cups, whereas synthetic smells were generated directly by \name, worn around the participant's neck. Participants could request repeated presentations of any physical or generated smell throughout the study. All 7-point ratings were collected verbally, with participants stating their rating aloud for the experimenter to log it via the \name\ interface. Optional natural language feedbacks were also collected verbally from the participants using the same interface. The set-up, odor presentation and interfaces used in this study were all shown in Figure~\ref{fig:user-study-procedure}.

Each participant completed 12 trials, with each trial consisting of three stages: 
(1) a 3-AFC olfactory discrimination task, (2) a direct perceptual similarity rating, 
and (3) optional natural-language refinement in the \name\ and Expert conditions.

\textit{3-AFC olfactory discrimination.} At the beginning of each trial, we first established the target reference. We presented 
the physical target and asked the participant to identify it. Regardless of their response, 
we then revealed the target descriptor (e.g., ``orange'') and asked them to rate their 
familiarity with the target smell on a 7-point Likert scale. We revealed the target identity 
to provide participants with an explicit perceptual reference for subsequent comparisons, 
consistent with prior work showing that expectations about an attended odor can elicit 
odor-specific predictive representations that facilitate subsequent olfactory 
perception~\cite{zelano2011olfactory}. Participants were next sequentially presented with three candidates in randomized order without revealing their identities: (1) the generated target smell, (2) a generated distractor smell, and (3) a physical distractor smell. Participants selected the candidate that most closely resembled the previously presented target and rated their confidence on a 7-point Likert scale. A response was considered correct when the participant selected the generated target. A 10-second interval separated consecutive odor presentations.

\textit{Perceptual similarity.} Next, participants directly compared the physical target with the generated target, presented in randomized order, and rated their perceived similarity on a 7-point Likert scale. A 10-second interval separated the two presentations.

\textit{Natural-language refinement.} Lastly, in the \name\ and Expert conditions, participants were invited to provide natural-language feedback describing how the generated target could better approximate the physical target. Their spoken feedback was transcribed and passed verbatim to \name\ together with the target descriptor and current composition. \name\ then generated a revised composition, which participants compared against the physical target and rated again for perceived similarity. Participants could skip refinement or provide multiple rounds of feedback, stopping when they were satisfied with the generated smell or did not feel confident that further feedback would improve it. The \name\ zero-shot condition did not include this refinement stage.

This procedure was repeated across all 12 trials. To mitigate olfactory fatigue, participants could take breaks at any time and were provided with a handheld fan to disperse residual odors between presentations. Participants were also encouraged to briefly smell their own skin or clothing if they experienced diminished olfactory sensitivity.

\subsubsection{Post-Study Interview}

After completing all 12 trials, participants took part in a semi-structured interview about their overall experience with \name, including generated smells they perceived as particularly similar to or different from their physical counterparts. Follow-up questions were asked where appropriate to clarify responses and elicit additional details. Participant responses were documented through typed notes.

\subsubsection{Analysis}

We analyzed three primary outcomes: 3-AFC accuracy, perceived similarity, and changes in perceived similarity following natural-language feedback. Because each participant evaluated 12 targets, we used mixed-effects models with crossed random intercepts for participant and target to account for repeated observations across both participants and smells.

3-AFC accuracy was binary and was therefore analyzed using a binomial generalized linear mixed-effects model (GLMM). Perceived similarity and confidence ratings were analyzed using linear mixed-effects models (LMMs). Fixed-effect $p$-values for LMMs were computed using Satterthwaite's approximation for degrees of freedom. Between-condition and pre--post feedback contrasts were estimated using estimated marginal means with Tukey adjustment for multiple comparisons.

Within-condition identification against chance was assessed with Wald $z$-tests on intercept-only mixed-effects logistic models of 3-AFC correctness, comparing the intercept to $\mathrm{logit}(1/3)$. Those tests were run overall and separately for \name\, Expert, and \name\ zero-shot.

Between-condition differences were evaluated in two models. First, a mixed-effects logistic model predicted correctness from Condition; Tukey-adjusted Wald contrasts compared \name, Expert, and \name\ zero-shot. Second, a linear mixed model predicted the 1--7 similarity rating collected after identification and before any spoken feedback from Condition; Tukey-adjusted $t$-tests compared the same three conditions.

Within-condition changes after spoken feedback were assessed only in \name\ and Expert, the two conditions with a feedback stage. A linear mixed model predicted similarity from Time (before vs.\ after feedback) $\times$ Condition. Simple Time effects within each condition tested the before--after gain. The Time $\times$ Condition interaction tested whether those gains differed. Trials without feedback retained the first rating after the feedback stage, so they contribute a zero change. The same Time $\times$ Condition model was then fit on the subset of trials where spoken feedback was given.

To examine whether smell family influenced identification, mixed-effects logistic models with and without cluster as a fixed effect, both controlling for condition, were compared with a likelihood-ratio test. The influence of pre-trial familiarity was tested by adding familiarity as a covariate to the identification model (Wald $z$-test). Pearson correlations further examined the relationship between familiarity and identification accuracy and between familiarity and the before-feedback similarity rating. A Pearson correlation between trial-level correctness and before-feedback similarity tested whether correct trials also felt closer to the target. Whether correct identification predicted participant confidence was tested with a linear mixed model of confidence on correctness, controlling for condition.

\subsection{Comparative User Study Results}
\label{sec:results}

\subsubsection{RQ1: People Identified Generated Smells Above Chance}

Participants successfully identified the intended reconstruction above chance, across 50 everyday targets spanning across 12 categories of smell. Wald $z$-tests from a mixed-effects logistic model compared 3-AFC accuracy to chance of one in three. Overall, 490 of 720 trials were correct (68.1\%; $z = 11.76$, $p < .001$). Each condition was above chance and produced statistically significant results: \name\ 70.8\% (170/240; $z = 8.00$, $p < .001$), Expert 69.6\% (167/240; $z = 7.86$, $p < .001$), and \name\ zero-shot 63.7\% (153/240; $z = 7.16$, $p < .001$).

While users often associate the synthetic smells generated by \name\ with the real target smell, we do not claim that the generated smell was perceptually indistinguishable from the target. The 12-channel system was limited to a fixed set of essential oil base odorants, so it could only approximate a real smell. What the result showed was that untrained participants could reliably identify \name-generated recreation of a real target smell amongst two unrelated distractors above chance across the 50 benchmark smells.

\subsubsection{RQ2: \name\ Led on Both Accuracy and Initial Similarity, but Condition Differences Were Not Significant}

On the 3-AFC test accuracy, \name\ scored the highest at 70.8\% (170/240): 1.2 points above Expert (69.6\%; 167/240) and 7.1 points above \name\ zero-shot (63.7\%; 153/240). Tukey-adjusted Wald contrasts from a mixed-effects logistic model found no significant pairwise differences: \name\ vs.\ Expert ($z = 0.28$, $p = .96$), \name\ vs.\ \name\ zero-shot ($z = 1.56$, $p = .26$), Expert vs.\ \name\ zero-shot ($z = 1.29$, $p = .40$). 

The same pattern held for perceived similarity, where before any feedback, \name\ was highest at 4.09: 0.30 points above Expert (3.79) and 0.32 points above \name\ zero-shot (3.77). Expert and \name\ zero-shot were nearly indistinguishable on the initial similarity rating. Tukey-adjusted $t$-tests from a linear mixed model found no significant differences: \name\ vs.\ Expert ($t = 1.39$, $p = .35$), \name\ vs.\ \name\ zero-shot ($t = 1.59$, $p = .26$), Expert vs.\ \name\ zero-shot ($t = 0.21$, $p = .98$). Standard deviations within the three conditions were close (\name\ $SD = 1.53$; Expert $SD = 1.46$; \name\ zero-shot $SD = 1.50$).

\begin{figure}[t]
    \centering
    \includegraphics[width=\linewidth]{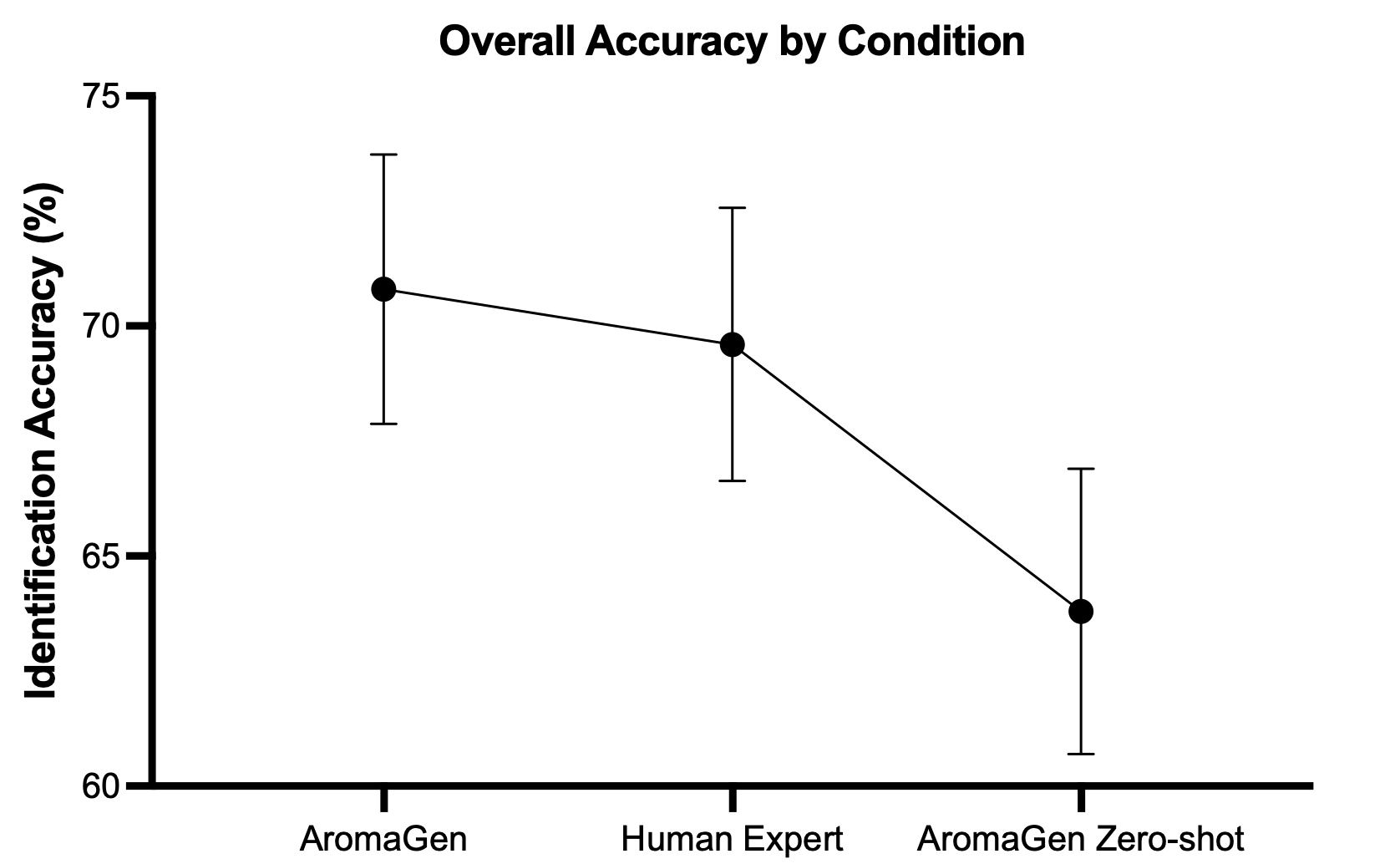}
    \caption{3-AFC test accuracy by composition condition ($N=60$; 240 trials per condition). Accuracy was above chance across all three conditions (\name: 70.8\%; Expert: 69.6\%; \name\ zero-shot: 63.7\%), with no statistically significant differences between conditions.}
    \label{fig:accuracy-condition}
\end{figure}

In short, the three conditions performed similarly, as shown in Figure~\ref{fig:accuracy-condition}. 
%We did not find a statistically significant result that adding RAG improved 3-AFC accuracy or perceived similarity compared with Expert or \name\ zero-shot (Figure~\ref{fig:accuracy-condition}). The means hide a modest tail difference.
We did however find that \name\ slightly outperformed on the initial similarity rating compared to the other two, as it had more high scores (47 of 240 trials at 6--7, 19.6\%) and fewer low scores (41 of 240 at 1--2, 17.1\%) than Expert (30 of 240 at 6--7, 12.5\%; 53 of 240 at 1--2, 22.1\%) or \name\ zero-shot (31 of 240 at 6--7, 12.9\%; 48 of 240 at 1--2, 20.0\%).

This result should not be interpreted as establishing equivalence between AI and Human-expert composition. Rather, within the scope of our 50 benchmark smells and fixed 12 base odorants, we found no evidence that replacing manually curated Human-expert compositions with AI-generated compositions reduced participants' ability to associate the synthetic generated smell to its intended target. Thus, the value of AI-based aroma composition may lie not in outperforming trained experts, but in making the ability to compose recognizable target aromas substantially more accessible, responsive, and scalable.

\subsubsection{RQ3: Effect of Spoken Feedback on Perceived Similarity}

After the similarity rating, participants in the \name\ and Expert conditions could optionally tell the system in their own words how they wanted to improve the synthetic smell. The device then played the updated smell, and they rated similarity again. Participants in the \name\ zero-shot condition skipped the feedback stage. A summary of the changes in the similarity ratings before and after feedback is shown in Figure~\ref{fig:cluster-accuracy}. 

\textit{\name.} Participants gave spoken feedback on 163 of 240 trials (68\%). Averaged across all 240 \name\ trials, mean similarity rose from 4.09 to 4.65 ($+0.56$, $t = 4.99$, $p < .001$); this average included the 77 trials where no feedback was given, whose ratings stayed unchanged. Those 77 trials had started at an average similarity rating of 5.35, which was quite high on the similarity scale such that participants could have chosen to skip improvement. For the 163 trials where feedback was given, the starting similarity was lower (3.50), and the gain after the revised mix was larger: up to 4.32 ($+0.82$, $t = 6.33$, $p < .001$).

\textit{Expert.} Participants gave feedback on 115 of 240 Expert trials (48\%). Averaged across all 240 Expert trials, similarity rose from 3.79 to 4.05 ($+0.26$, $t = 2.31$, $p = .021$). For the 115 trials where participants gave feedback, similarity rose from 3.48 to 4.02 ($+0.54$, $t = 3.49$, $p < .001$); for the 125 where they did not, similarity stayed at an average of 4.08. Nearly all feedback provided in both conditions were a single round.

We cannot conclude that spoken feedback helped \name\ composition more than human-expert composition ($t = -1.90$, $p = .058$). What differed is how close the smell already felt, and how often people asked to change it. Before any feedback, participants rated \name\ compositions closer to the real object (mean 4.09 vs.\ 3.79) and requested a revision more often (68\% vs. 48\%). They skipped feedback when the smell already felt close (mean 5.35 vs.\ 4.08). Across both conditions they asked for a change when it felt far off (3.5 on the 7-point scale).

\begin{figure}
    \centering
    \includegraphics[width=\linewidth]{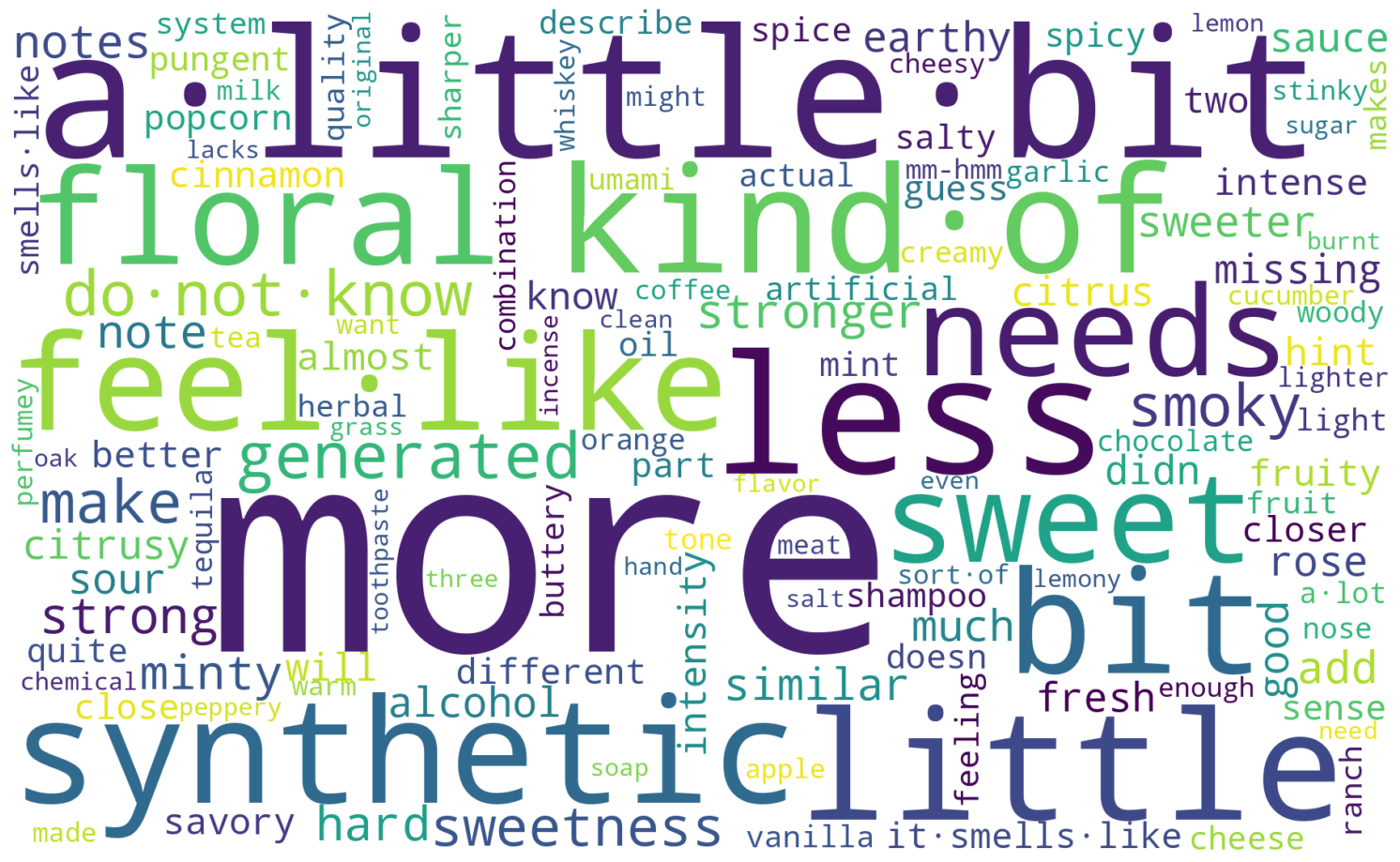}
    \caption{Phrase cloud of participants' natural-language feedback. Feedback commonly expressed directional changes to perceptual qualities ("more", "less" etc), relational comparisons to desired smells ("similar", "more like" etc), and uncertainty or approximation ("don't know", "maybe" etc), illustrating how participants could steer generated smells without explicitly specifying odorants or compositional changes.}
    \label{fig:word cloud}
\end{figure}

\subsubsection{Qualitative analysis of feedback patterns}
The way participants provided their feedback also revealed several recurring patterns in how they attempted to refine generated smells. A phrase cloud of all participants' feedback is shown in Figure~\ref{fig:word cloud}. Feedback frequently specified particular perceptual qualities to strengthen or reduce, including floral, sweet, smoky, earthy, minty, citrus, and sour characteristics, and most notably, participants used additive or intensifying language substantially more often than reductive language, with \textit{more} appearing 278 times compared with 60 occurrences of \textit{less}. Feedback almost never exclusively contained negative advice, except for a few cases such as P12's feedback on the generated black pepper smell: "they should smell less like cheddar". Rather, most feedback simultaneously included changes in both directions, such as P6's feedback on \name's generated cucumber smell: "I would say a little bit more grassy and less floral".

Participants also commonly described the generated aromas relationally rather than in terms of precise composition, using expressions such as ``more like,'' ``closer,'' and ``similar,'' suggesting that refinement was often also framed as steering the perceived smell toward an experiential target in addition to explicitly manipulating constituent perceptual qualities. This could be seen in P1's feedback on the synthetic fries with ranch sauce smell: "the smell should be more oily, like fried food, and also a little bit more like buttermilk". 

At the same time, participants' feedback contained frequent markers of uncertainty and approximation, including terms such as \textit{don't know}, \textit{maybe} and \textit{like}, suggesting that participants were not always able to precisely articulate their desired olfactory changes, even if they detected a mismatch in the smell. This could be seen in P12's feedback on the synthetic basil smell: "so it's very soft, I would increase the intensity and make it smell more like a grass field. Um I don't know how to say, like it's missing some earthy tones".

Together, these observations suggest that participants could communicate directional perceptual changes through natural language even when they had difficulty articulating the perceived mismatch. Their feedback therefore did not require explicit identification of constituent odorants or specification of compositional changes, but could instead be expressed through approximate perceptual qualities, comparisons, and desired directions of change.

\subsubsection{Across Category Analyses}
\label{sec:secondary}

\begin{figure*}[t]
    \centering
    \includegraphics[width=\linewidth]{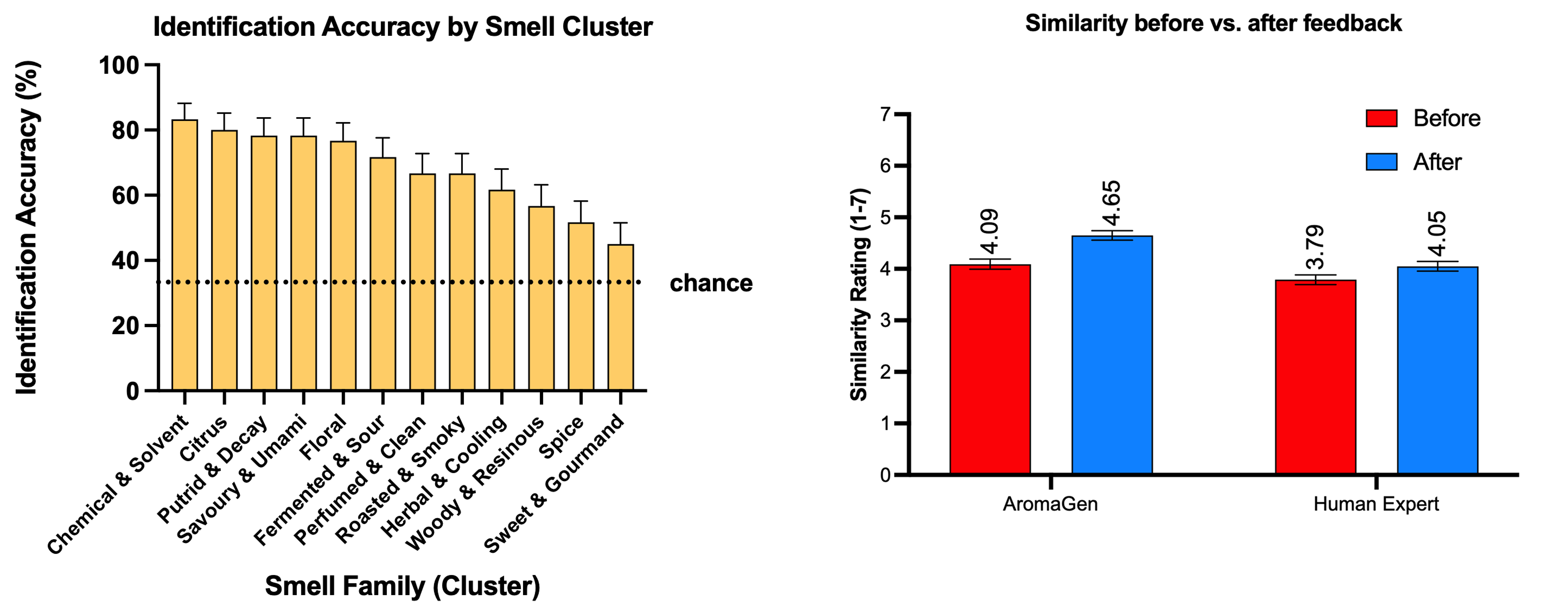}
    \caption{Three-alternative forced-choice identification accuracy by olfactory cluster, pooled across conditions (N = 60, 720 trials). (Left) Identification accuracy by smell cluster; all 12 clusters exceed chance (33\%), ranging from 83\% (Chemical \& Solvent) to 45\% (Sweet \& Gourmand). (Right) Similarity ratings before and after spoken feedback for \name{} and Human Expert; both conditions show a significant increase in perceived similarity following feedback (\name: 4.09→4.65; Human Expert: 3.79→4.05).}
    \label{fig:cluster-accuracy}
\end{figure*}

Some categories of smell were much easier to identify than others. A likelihood-ratio test on mixed-effects logistic models with and without cluster found that smell family predicted identification ($\chi^2(11) = 30.19$, $p = .001$). Chemical and solvent smells (83\% correct), citrus (80\%), putrid and decay (78\%), and savoury and umami (78\%) were the easiest. Sweet and gourmand (45\%), spice (52\%), and woody and resinous (57\%) were the hardest (Figure~\ref{fig:cluster-accuracy}). The easiest individual targets were mint fluoride mouthwash, identified correctly by all 17 participants who encountered it, and peony-and-rose shampoo (16/17). The hardest were cinnamon roll (4/14) and incense (4/13). The harder clusters tend to involve smells that are complex mixtures or whose character does not map cleanly onto any single cartridge. To better understand how the harder clusters were often confused with one another, we analyzed the smell family options selected on incorrect trials. We used a confusion matrix to tabulate, for every pair of smell families, how often an option from one family was chosen when it appeared against a target from the other, normalized by how often that pairing was presented. Against a base rate of 16\% across all distractor presentations, the most confusing pairs were Citrus with Perfumed and Clean (41.9\%), Sweet and Gourmand with Woody and Resinous (37.8\%), Putrid and Decay with Roasted and Smoky (35.0\%), and Fermented and Sour with Spice (33.3\%). The most reliably distinguished pairs were perceptually opposed and fell well below that rate: Citrus with Savoury and Umami (2.9\%), Floral with Putrid and Decay (4.0\%), and Herbal and Cooling with Savoury and Umami (4.4\%).

\textit{Familiarity.} Participants' self-reported familiarity with the target smell was not associated with either 3-AFC accuracy ($r = .00$, $p = .94$) or perceived similarity ($r = -.03$, $p = .38$). A mixed-effects logistic model likewise found no significant effect of familiarity on 3-AFC accuracy ($z = 0.27$, $p = .79$), and including familiarity as a covariate did not alter the observed condition comparisons. Thus, greater self-reported familiarity with a target smell did not confer an advantage in distinguishing its \name-generated reconstruction. We will return to this finding in the Discussion.

\textit{Confidence and accuracy.} When participants correctly identified a smell, they were 0.70 points more confident on the 7-point scale ($t = 5.81$, $p < .001$). They also gave the correctly identified smell a higher similarity rating on average (4.06 vs.\ 3.51 when incorrect; Pearson $r = .17$, $p < .001$), so the trials where the smell was recognizable were also the trials where it felt closer to the real thing.

\subsubsection{Compositional Analyses}

\begin{figure}[t]
    \centering
    \includegraphics[width=\linewidth]{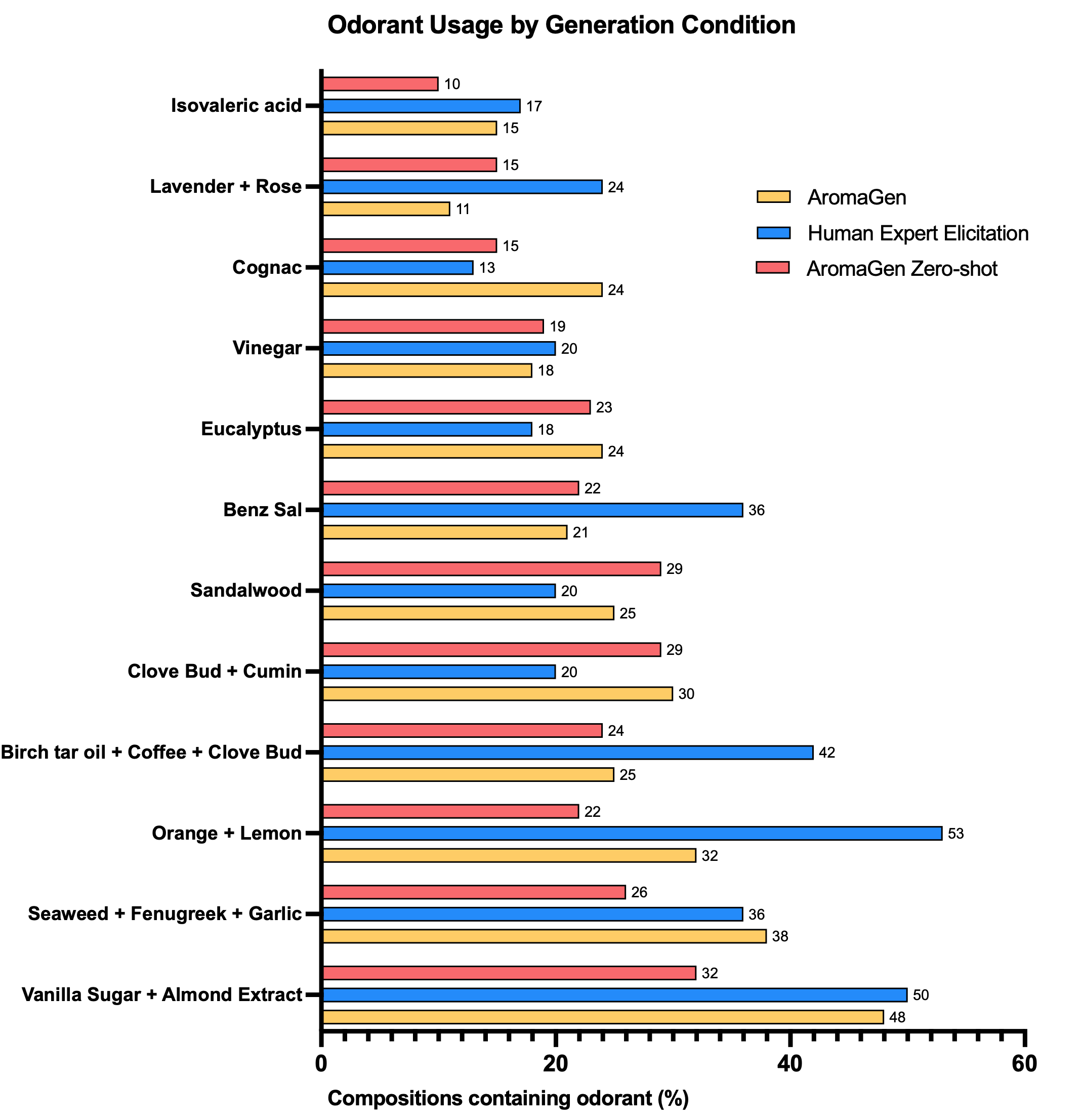}
    \caption{Percentage of compositions containing each of the 12 base odorants, by generation condition (\name, Human Expert Elicitation, \name{} Zero-shot). Usage patterns differ notably by condition; Human-expert compositions rely most heavily on a small number of signature odorants (e.g., Orange + Lemon, 53\%), while \name{} Zero-shot spreads usage most evenly across the palette}
    \label{fig:compositional analysis}
\end{figure}

We compared the composition of the generated ratios for each target smell across the three study conditions: \name{}, Expert Elicitation, and \name\ zero-shot. The composition text was parsed into shared ingredients over the study's fixed 12-odorant base palette, using the composition each participant got. All target smells were present in all three conditions, and for each of these targets we computed the mean ingredient-share vector per condition and quantified divergence using total variation distance (TVD) to produce a score (0--100), where 0 means the two conditions used identical ingredients in identical proportions and 100 means they shared no ingredients. This analysis showed that both AromaGen and \name\ zero-shot diverged from the Human-expert composition by a similar overall magnitude ({TVD: AromaGen and Expert = 40.5; AromaGen zero-shot and Expert = 41.6}), indicating that AromaGen and \name\ zero-shot have different composing strategies than a human expert. However, some smells showed strong convergence across all three conditions ({Mint chewing gum, peony and rose oil shampoo, Bacon, and Lavender}), while others diverged in both AI conditions: for {Black pepper}, Expert used Clove Bud + Cumin at just {14\%} versus 74\% (AromaGen) and 77\% (AromaGen zero-shot); for {Cucumber} and {Aloe vera}, Expert leaned on Benz Sal ({81\% and 58\%}, respectively), where both AI conditions used far less (2--32\%) or none; for  {Pickled cucumber}, Benz Sal was used ({48\% in Expert versus 0\% in both AI conditions}); and for  {Nacho with sour cream}, Expert centered on Vinegar ({44\%}) where AromaGen used only 7\%, leaning instead on Seaweed + Fenugreek + Garlic ({30\%}, Expert's own second-largest ingredient at 38\%) and Clove Bud + Cumin ({25\%}, versus 7\% in Expert). The cumulative odorant usage of 12 base odorants across all three conditions is shown in Figure~\ref{fig:compositional analysis}.

\subsection{Exploratory Free-form Aroma Generation Study}

Our comparative study evaluated \name\ against a controlled benchmark of predefined real-world smells, but did not capture how users might engage with the system when free to determine what they wanted to smell. We therefore conducted a small-scale exploratory study ($n=5$) in which participants could freely request and iteratively refine olfactory experiences through natural language. Rather than statistically evaluating system performance, this study sought to identify preliminary user behaviors, strength, limitations and perceived affordances of \name, and opportunities for creative olfactory exploration beyond the predefined benchmark. All participants used the AromaGen version of \name\ for their free-form aroma generation in this study.

\subsubsection{Participants}

We recruited five participants from the participant pools of our preceding expert elicitation and comparative user studies. We intentionally recruited participants with prior experience using \name\ so that they entered the study with an existing understanding of the system and its interaction mechanisms. Because our goal was to explore the creative possibilities of free-form aroma generation, we further prioritized participants whose professional backgrounds involved creative practice or expertise in the composition and perception of smells.

The recruited participants included a perfume hobbyist, a taste researcher, a pastry chef, a sommelier, and a cooking enthusiast. Their prior experience with \name\ and relevant professional backgrounds allowed the study to focus on how experienced users explored the system's capabilities rather than on initial system familiarization. Participants were compensated at a rate of \$25 per hour.

\subsubsection{Procedure}

Each participant completed an approximately one-hour session using \name\ with the same set of 12 base-odorant cartridges as in the comparative study. We first introduced \name\ as a system capable of generating any olfactory experiences from free-form natural-language descriptions and refining them through natural-language feedback. Participants then completed two standardized onboarding tasks: (1) recreating the smell of the last meal they ate, and (2) creating a smell representing a memorable experience or moment from their life. These tasks familiarized participants with generating both smells from concrete, physical sources and abstract and experiential olfactory experiences.

For each request, \name\ generated an initial composition, which the participant smelled and rated on a 7-point Likert scale in response to the question, ``How well does this smell match your expectation?'' Participants could then engage in the same feedback procedure as in the AromaGen and expert conditions in the comparative study to iteratively refine the smell. No fixed limit was imposed on the number of refinement iterations, although participants were reminded of the remaining study time and their current iteration count after the third iteration of a smell.

Following onboarding, participants were given the remainder of the session to freely create 10 smells of their choice. We intentionally did not ask participants to prepare a list of smells before the study, allowing their choices to develop through interaction with \name. Requests could involve concrete objects, foods, fragrances, places, memories, experiences, or abstract and imagined smells. Participants could iteratively refine each generation through natural-language feedback and provided a 7-point match rating after each generated composition. We logged all prompts, feedback, generated odorant compositions, ratings, and iteration histories.

\subsubsection{Post-Study Interview}

After the free-form exploration, we conducted a semi-structured interview examining three aspects of participants' experiences: \name's perceived capabilities and failure cases, changes in their understanding of what the system could create, and how open-ended interaction supported creative exploration. We asked participants to reflect on their most and least successful generations, the kinds of smells they believed \name\ could or could not create, whether these expectations changed through use, whether interaction inspired new olfactory ideas, and how they might use such a system in everyday work, study, or other activities.
Participants' responses were documented and typed into their corresponding fields in the data collection sheets.

\begin{figure}
    \centering
    \includegraphics[width=\linewidth]{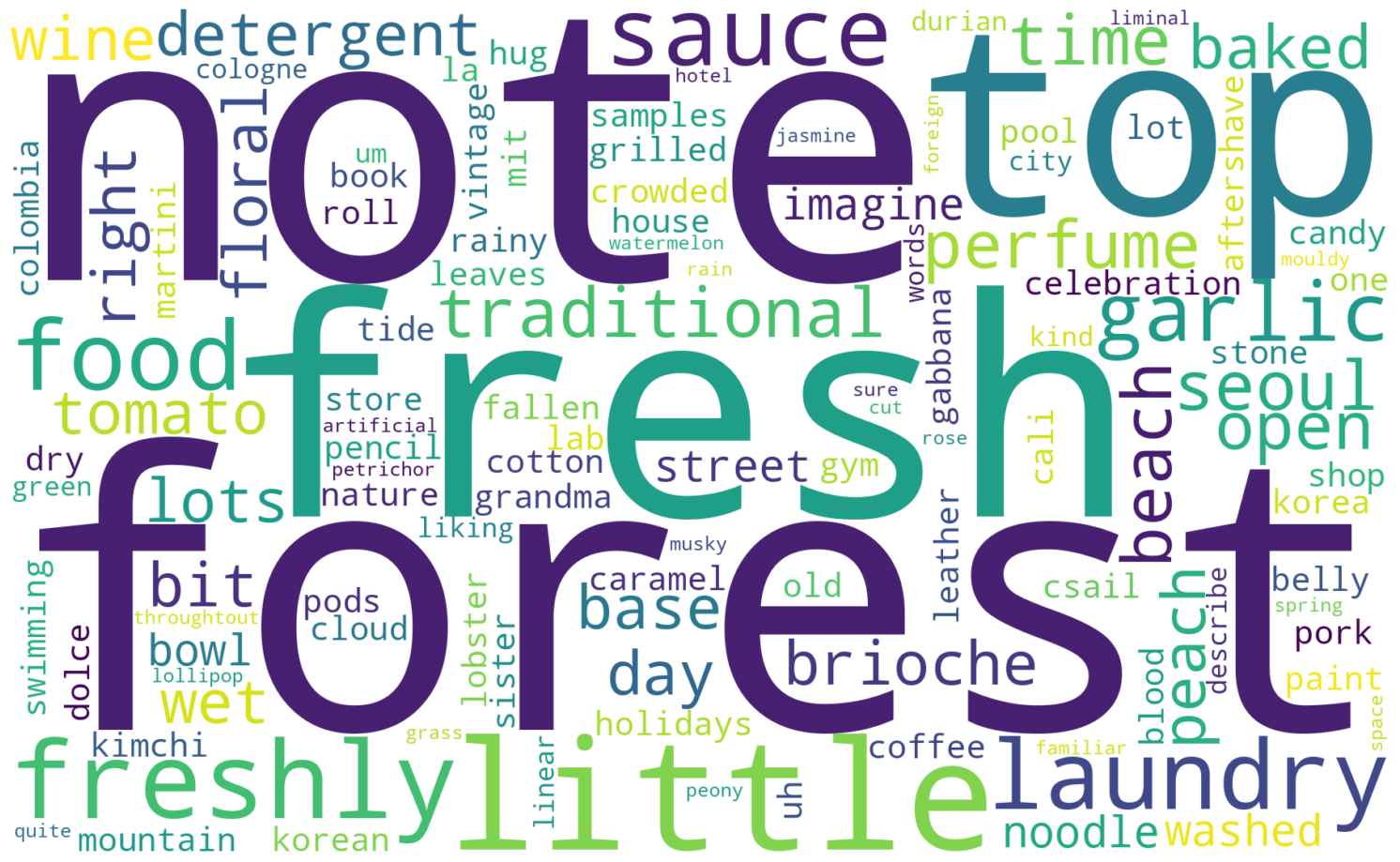}
    \caption{Word cloud of participants’ freely generated target aroma, spanning nature ("forest", "mountain" etc), edibles ("kimchi", "sauce" etc), chemicals ("laundry", "detergent" etc.), places ("Colombia", "Seoul" etc), people ("grandma", "sister" etc) and more.}
    \label{fig:free form phrase cloud}
\end{figure}

\subsubsection{Preliminary Findings from Free-form Aroma Generation}

Across the 50 free-form aroma requests, initial perceived-similarity ratings ranged from 1 to 7 (\(M=3.42, SD=1.51\)). Participants requested at least one round of refinement for 32 of the 50 targets (64\%), while 18 targets received no further refinement. Among the 32 refined targets, perceived similarity increased from \(M=2.94\) (\(SD=1.05\)) for the initial generation to \(M=4.09\) (\(SD=1.23\)) for the final generation, corresponding to a mean improvement of 1.16 points, again demonstrating the effectiveness of natural language based feedback for synthetic smell improvement. Twenty-four of these 32 targets (75\%) ended with a higher similarity rating than their initial generation, while eight remained unchanged and none ended with a lower rating after feedback. Across all 50 targets, the mean final similarity rating was \(4.16\) (\(SD=1.46\)). Overall, our preliminary results demonstrated that \name's smell composition pipeline could reasonably reconstruct participant's freeform aroma requests, though at a lower fidelity compared to its generation of our 50 benchmark smells.

We analyzed the post-study interviews using qualitative content analysis~\cite{graneheim2004qualitative}, focusing primarily on the manifest content of participants' responses. Given the exploratory nature and small sample ($N=5$) of this study, we used the resulting descriptive categories to organize preliminary observations rather than to make claims of generalizability beyond the participants studied.

Overall, participants explored a diverse range of requests, spanning foods, fragrances, natural environments, personal memories, and abstract or imagined olfactory experiences. Figure~\ref{fig:free form phrase cloud} summarizes participants' initial descriptions of their 10 favorite smells. Our analysis identified four preliminary observations for further investigation. 

\textbf{Strengths and limitations in smell reconstruction.} Reconstruction quality varied substantially across the 50 free-form targets, with initial similarity ratings spanning the full 1--7 scale. Several targets received high initial ratings, including ``Sister's hug,'' ``Spring breeze laundry detergent with the scent of warm laundry,'' and ``freshly washed laundry'' (all 7/7), whereas targets such as ``the beach'' (1/7) and ``catharsis,'' ``Open air Chinese wet market,'' and seafood pasta (2/7) were rated substantially lower. 

Participants' interview responses provided additional context for these differences, where they identified several types of targets that \name\ reconstructed relatively well, alongside others that remained difficult. Some preliminary commonalities emerged around smells with distinctive or readily describable characteristics. P1, for example, found relatively ``monotonic'' smells associated with a clear olfactory description, such as the woody character of a forest, easier to reconstruct than compositionally complex smells such as kimchi. P4 identified artificial scents such as lollipops and laundry detergent among her most successful generations, while P5 similarly found laundry and fresh, fruity smells successful. Conversely, participants identified difficulties with complex compositions, subtle environmental smells, and olfactory characteristics poorly represented by the available basis: P5's seafood pasta, for example, remained notably lacking in fishiness despite repeated attempts.

However, these observations were not uniform and occasionally contradicted one another. P2 considered a generated forest smell particularly successful and felt that \name\ reproduced some natural elements well, whereas P5 specifically identified faint natural smells such as a forest after rain as difficult to reproduce. P2 found chemical targets such as paint difficult, while P4 and P5 considered several artificial or synthetic smells successful. Rather than suggesting a simple distinction between categories that \name\ can and cannot reconstruct, these preliminary observations indicate that reconstruction quality may depend on an interaction among the target's olfactory distinctiveness and complexity, degree of artificiality, the available odorant basis, and participants' individual expectations and associations.

\textbf{Perceived affordances and boundaries of \name.}
Participants developed different mental models of \name's generative capabilities through interaction. P1 reasoned that targets associated with clear linguistic descriptions should be easier for \name\ to generate, whereas objects or experiences that are difficult to characterize through smell-related language may present a greater challenge. P4 instead described the system as analogous to a ``printer'' for smells, whose generative possibilities ultimately depend on the essential oils available in its palette. P3 initially believed that iterative refinement could allow \name\ to create essentially any smell, but her subsequent experiences with unfamiliar and imagined targets led her to recognize potential boundaries to this capability. 

These accounts preliminarily suggest that participants understood \name's affordances through multiple aspects of the system: its natural language-based input, its finite physical odorant basis, and their experiences of successful and unsuccessful generations. Importantly, these interpretations varied across participants and should be treated as emerging mental models rather than stable properties of how users understand the system.

\textbf{Creative exploration beyond existing smells.}
For some participants, experiencing \name's outputs prompted them to move beyond reconstructing known smells toward manipulating, imagining, or inventing olfactory experiences. P2, after finding a generated forest smell similar to their intended target, began considering variations of the same environment across weather and seasons, such as a forest in winter or after rain. P4 described the interaction as `fun and playful'' and explored unconventional targets such as the smell of "catharsis". Most notably, P3's exploration developed from a personally meaningful abstract target to an imagined one. After experiencing a successful generation of `Sister's hug,'' she subsequently attempted to generate the smell of a cloud, an olfactory target without a conventional real-world referent, and described feeling that the combination of her imagination and the system gave her ``endless new ideas.''

These cases illustrate a potential generative role for \name\ in which experiencing an AI-generated smell can itself prompt subsequent olfactory ideation, leading users to explore abstract, imagined, or unconventional targets that they had not initially considered. Given the small sample, we treat these cases as preliminary evidence of a possible creative affordance rather than evidence that \name\ generally increases olfactory creativity.

\textbf{Personal memories and experiences.}
Participants also used \name\ to construct smells associated with autobiographical memories and personally meaningful experiences. These targets could be difficult to reconstruct because their intended olfactory qualities were often implicit and highly individualized. P1, for example, found requests such as ``my grandma's house'' difficult to communicate through feedback and suggested that the system could instead ask contextual questions about the foods, people, season, and circumstances associated with the memory. At the same time, approximate olfactory representations could still become emotionally meaningful when connected to an existing personal association. P5 imagined using the system to recall entering his grandmother's kitchen or to allow others to experience aspects of his food culture.

One particularly salient instance occurred when P3 generated `Sister's hug,'' a floral and sweet smell intended to represent the experience of hugging her sister. While smelling the generated aroma, P3 had tears welling in her eyes and described experiencing excitement, curiosity, and homesickness. She explained that the smell `bring[s] her into a moment that she really wants'' and reminded her of her sister. She subsequently envisioned using \name\ when she missed home or wanted to recreate childhood memories. While this individual episode cannot establish that \name\ reliably evokes autobiographical memories or emotional responses, it illustrates how even an approximate synthetic smell may acquire personal and emotional significance through a user's pre-existing associations with the imagined experience.

Overall, these preliminary findings suggest several directions for future investigation rather than definitive conclusions about \name's free-form capabilities. In particular, they motivate examining how olfactory distinctiveness, compositional complexity, linguistic describability, available base odorants, and individual expectations jointly influence reconstruction quality, how users form mental models of the capabilities and boundaries of language-driven olfactory generation, and whether interaction with generated smells can support creative exploration or personally meaningful experiences. These observations should be interpreted in light of the small sample size and participants' prior experience with \name.

\section{Discussion}

\subsection{Human--LLM Collaboration for Olfactory Creation}

Our findings suggest a role for generative AI in olfactory creation that extends beyond automating the work of a perfumer. Rather, \name\ enables a division of labor between human and machine that reflects an asymmetry in human olfactory cognition: generating an appropriate composition from constituent odorants is difficult, whereas evaluating whether an encountered smell corresponds to an intended target can be considerably more accessible, even for novice users.

Creating smells from constituent materials traditionally requires substantial learned expertise. Professional perfumery involves developing knowledge of individual materials, their interactions, and their perceptual effects in mixtures ~\cite{holley2002cognitive}.
Olfactory expertise is also associated with differences in odor memory and imagery; professional perfumers can evoke olfactory images more readily than novices, whereas olfactory imagery in the general population is difficult to evoke and highly variable ~\cite{royet2013impact}. Composition is further complicated by the configural perception of odor mixtures: as components are combined, they may be perceived as an integrated odor object whose constituents are difficult to identify independently ~\cite{thomasdanguin2014perception}. Consequently, a novice attempting to construct a target smell faces minimally two coupled problems: determining which constituent materials and proportions could produce the target, and maintaining
a sufficiently useful and stable olfactory mental imagery of the desired smell to guide the recreation of that target.

Yet, difficulty in composing or explicitly identifying a smell does not imply an equivalent inability to evaluate one. Humans exhibit strong olfactory discrimination abilities despite considerable difficulty freely identifying and
naming odors ~\cite{dewijk1994odor,royet2013impact}, and expectations about an odor can elicit predictive representations that facilitate subsequent perception~\cite{zelano2011olfactory}. Together, these findings suggest an asymmetry particularly relevant to interactive aroma generation: users may struggle to
explicitly identify, decompose, or construct a desired smell while remaining capable of evaluating candidate smells against an expected target.

\name\ operationalizes this asymmetry as a human--LLM collaboration. The LLM translates users' descriptions into compositions over the 12-odorant basis, while users smell the resulting candidates and communicate perceptual discrepancies through natural language. Instead of specifying constituent odorants and ratios, users can express that a generation should be sweeter, less smoky, or more citrus-like, leaving \name\ to translate these perceptual directions into compositional changes. \name\ thus shifts the user's role from \textit{constructing a solution} to \textit{evaluating and steering candidate solutions}: the model navigates a compositional space that users may struggle to parameterize directly, while users retain perceptual judgment over whether the resulting experience succeeds. Importantly, shifting this compositional burden to the LLM does more than redistribute work between human and machine; it may also lower the expertise and effort required for people to create and personalize olfactory experiences themselves.

\subsection{Toward Accessible Personalized Aroma Generation}  This shift points toward a broader opportunity for making personalized aroma creation more accessible. Conventional formulation requires not only compositional expertise, but also the manual effort of selecting, measuring, mixing, and iteratively adjusting constituent materials. During an early internal iteration of our expert elicitation procedure, an expert manually formulated target smells by pipetting and mixing individual odorants before comparing each formulation with its physical target. Within a one-hour session, the expert completed formulations for only nine targets. Although this observation was not part of our controlled evaluation and should be interpreted only as an illustrative case, it highlights the material and temporal overhead that can accompany manual formulation.  

\name\ changes where this effort occurs. Rather than requiring users to manipulate constituent materials directly, users can begin with an open-ended description of the olfactory experience they want, receive a physically rendered candidate, and personalize it through descriptions of how the resulting smell should change. In this way, the same natural language used to express an olfactory intention also becomes a means of iteratively steering its physical realization. This interaction could make aroma creation accessible to people who possess neither the materials nor the compositional expertise traditionally required to formulate smells, while enabling rapid exploration and personalization of different olfactory experiences.  

More broadly, this suggests an opportunity for AI-enabled sensory interfaces in which generative models handle aspects of sensory creation that are difficult for people to explicitly parameterize, while humans retain control through perceptual evaluation and subjective preference. For olfaction, this transforms generation from a one-shot prediction problem into an iterative process of \textit{describe, generate, smell, evaluate, and refine}. Rather than replacing human sensory expertise or judgment, systems such as \name\ could make the ability to create and personalize sensory experiences available to a broader range of users while keeping human perception at the center of the interaction.

\subsection{Familiarity Is Not the Same as Knowing a Smell}
\label{sec:fam-dis}

The preceding discussion highlights the central role of human perceptual judgment in interactive aroma generation. This raises a related question: how does users' prior experience with a smell shape their ability to evaluate its generated reconstruction? Surprisingly, participants' self-reported familiarity with the target predicted neither 3-AFC performance nor perceived similarity. Greater familiarity with an olfactory concept therefore did not necessarily translate into an advantage when evaluating its \name-generated reconstruction.

Our observations during the study suggest that this result may reflect different ways in which participants ``knew'' a smell. Some participants immediately recognized a physical target and subsequently reported high familiarity. Others could not initially identify the target, but recognized it as something they had encountered frequently once its identity was revealed and likewise reported high familiarity. Still others remained uncertain about the physical odor even after seeing its label and consequently reported lower familiarity. Thus, similar familiarity ratings could reflect quite different relationships between prior exposure, spontaneous odor recognition, and recognition prompted by a semantic label. Because familiarity was measured after participants encountered the physical target and its identity was revealed, our measure may therefore capture familiarity with the labeled odor concept rather than the ability to independently retrieve a precise olfactory representation from prior experience.

This distinction is particularly relevant to \name's evaluation. Participants did not need to retrieve the target smell solely from memory: immediately before making their 3-AFC judgment, they smelled the physical target and were explicitly given its identity. Their task was therefore to evaluate generated candidates against a recently established perceptual reference. Participants may nevertheless have approached this comparison differently. Some appeared to rely primarily on the immediately presented physical reference, whereas others drew more strongly on prior experiences and expectations of what the labeled target ``should'' smell like. Under these conditions, greater prior familiarity may provide little additional advantage once a concrete perceptual reference has already been established.

The absence of a familiarity effect should therefore not be interpreted as evidence that prior olfactory experience is irrelevant to evaluating generated smells. Rather, it highlights that self-reported familiarity alone may be a poor proxy for the perceptual knowledge users bring to olfactory interfaces. Future evaluations could distinguish prior exposure, spontaneous identification, label-cued recognition, and confidence in one's remembered odor representation. Doing so could provide a more precise account of how prior olfactory knowledge shapes people's ability to evaluate and steer generated sensory experiences.

\subsection{Limitations and Future Work}

\textit{Hardware and palette coverage.}
The current device renders a composition as a sequence of single-channel pulses; only one channel is active at any given time, so odorants are only physically blended in mid-air and after release. This is a fundamental constraint on how faithfully the device can approximate a target smell. Future hardware designs could incorporate a mixing chamber, or a multi-outlet wearable or handheld form factor, that releases two or more base odorants concurrently, reducing the reliance on temporal interleaving pulses to create a perceptual blend. A compact handheld system would also lower the barrier to everyday and field use compared with a neck-worn dispenser.

The 12 base odorants setup introduces a second physical ceiling. Some targets cannot be approximated regardless of how well the AI composes, because the required olfactory quality is simply absent from the available basis. Our study does not separate failures due to composition (the right bases exist but were weighted poorly) from failures due to coverage (the 12 base odorants cannot represent the target at all). Disentangling these two error types is a prerequisite for knowing whether improvement should come from better AI composition, a richer or more strategically chosen palette, or both. Expanding the palette beyond 12 base odorants or allowing cartridge swaps for specific task domains is a potential way to offset this physical limitation.

\textit{Verbal mediation in refinement.}
The current refinement loop requires participants to translate their olfactory experience into words. As \citet{majid2021} notes, English speakers rarely have dedicated smell terms, and naming even familiar aromas is surprisingly difficult. Free-form spoken feedback therefore captures an evaluation only to the degree that people can put that evaluation into language. Future refinement interfaces could reduce this demand by eliciting preference among candidate mixtures, rather than requiring an explicit verbal diagnosis. That design would still leave generation to the model and evaluation to the human, without asking users to name the mismatch.

\textit{Learning from perceptual feedback.}
The current system accumulates human evaluations as in-context retrieval examples. Compositional knowledge therefore grows in the retrieval database but does not update the underlying model weights. This means the model cannot improve its olfactory representations across sessions. Incorporating perceptual feedback as a training signal could treat similarity ratings and refinement trajectories as preference data for fine-tuning and allow the system to improve in ways that retrieval alone cannot. Our study provides 278 spoken refinement rounds and paired before/after similarity ratings that could seed such a training process. 

\textit{Study scope.}
Our 50-target smell benchmark spans 12 olfactory clusters but skews toward food and beverage sources that are practical to prepare and verify in a lab setting. Performance on abstract, environmental, or autobiographically grounded smells may differ substantially, as suggested by our free-form study. Participants were recruited primarily from two U.S. university communities, limiting generalization across cultural backgrounds and olfactory familiarity. Olfactory fatigue is also a genuine constraint: 12 trials per session is near the practical limit for rigorous smell comparison, and finer-grained measures of fatigue and session order effects were not collected. Finally, essential oils are imperfect proxies for real-world sources; the perceptual gap between a base odorant and the corresponding source material may impact both composition quality and the verbal feedback participants can provide.

\section{Conclusion}

We presented \name, an AI-powered 12-channel wearable olfactory system that translates free-form natural-language descriptions of desired olfactory experiences into physically rendered compositions over a compact 12-odorant basis, while supporting iterative refinement through natural-language feedback. Across our evaluation, participants were able to distinguish generated target smells above chance in a 3-AFC olfactory discrimination task. Natural language-based feedback further improved the perceived similarity of generated smells to their physical targets.

Our findings do not suggest that \name-generated smells are perceptually indistinguishable from their real-world counterparts, nor that \name\ composition outperforms human experts. Rather, they demonstrate that a pretrained multimodal LLM, grounded in specialized human aroma composition data and coupled with a physically realizable olfactory basis, can translate open-ended language into smells that people can meaningfully associate with intended real-world targets. More broadly, \name\ demonstrates an interaction paradigm in which AI handles low-level olfactory composition while users evaluate and guide the resulting perceptual experience through natural language feedback, enabling people without olfactory compositional expertise to generate and iteratively refine physical olfactory experiences. Looking forward, systems like \name\ could make digital olfactory experiences more accessible and scalable across domains such as entertainment, education, and healthcare, contributing toward a future in which smell, a fundamental yet comparatively understudied human sense, becomes a more expressive and accessible medium for human--computer interaction.

\section{GENERATIVE AI DISCLOSURE}
During the manuscript writing, generative AI tools were used to refine 
phrasing in author-drafted text. All generated or refined text was 
reviewed, revised, and approved by the authors prior to inclusion. The authors take full 
responsibility for the accuracy and integrity of all content in this paper. 
The use of a large language model as a component of the \name\ system 
is described in Section~\ref{sec:interaction} and is the subject of this study.

\bibliographystyle{ACM-Reference-Format}
\bibliography{refs}

% =========================================================
% APPENDIX
% =========================================================

\appendix
\section{Describable Olfactory Space Construction}
\label{app:olfactory-space}

We adapted the corpus-based approach of H{\"o}rberg et al.~\cite{horberg2022} to construct a semantic representation of English olfactory descriptions. We used the UMBC WebBase Corpus, comprising approximately 2.97 billion tokens, as the underlying language corpus. Following H{\"o}rberg et al., we extracted an olfactory subcorpus containing sentences associated with olfactory and gustatory seed terms (e.g., \emph{odor}, \emph{smell}, \emph{scent}, \emph{fragrance}, and \emph{aroma}), yielding 92,120 sentences.

\subsection{Olfactory Descriptor Extraction}

We identified candidate olfactory descriptors by extracting content words occurring within a $\pm4$-word window of these seed terms and retaining candidates occurring at least 10 times. For each candidate, we calculated its Olfactory Association Index (OAI), which measures how strongly the word is associated with olfactory contexts relative to its overall occurrence in the corpus. This process yielded 3,558 candidate descriptors.

\subsection{Semantic Representation and Dimensionality Reduction}

We trained a 300-dimensional CBOW Word2Vec model on the combined olfactory and gustatory subcorpora and represented each retained descriptor using its learned embedding. Pairwise semantic distances were calculated from Pearson correlations between descriptor embeddings as

\begin{equation}
    D_{ij} = 0.5(1-\rho_{ij}),
\end{equation}

producing a descriptor-by-descriptor semantic distance matrix. We standardized this matrix and applied principal component analysis (PCA) to identify the dominant dimensions along which English olfactory descriptors were organized.

\subsection{Robustness Across Olfactory Association Thresholds}

Rather than selecting a single OAI cutoff a priori, we repeated the analysis using the top 5\%, 10\%, 15\%, 20\%, 25\%, and 30\% of candidate descriptors ranked by OAI. This allowed us to examine whether the semantic dimensions remained stable as increasingly less olfactory-specific vocabulary was introduced.

We further evaluated component stability through 200 bootstrap iterations and interpreted the resulting components by correlating descriptor scores with external semantic variables, including valence, arousal, sensory-modality associations, concreteness, imageability, and sensory experience.

\subsection{Resulting Semantic Dimensions}

Across these OAI thresholds, two prominent and stable semantic dimensions emerged. The first primarily differentiated pleasant from unpleasant olfactory descriptions, while the second differentiated descriptors referring to concrete odor sources from more abstract or evaluative descriptions. Based on these analyses, we retained these two dimensions as an operational two-dimensional representation of \name's describable olfactory space.

\section{Expert Interview Protocol}
\label{app:expert_interview}

\textbf{Expert Background}
\begin{itemize}
    \item How many years of studies or practice do you have in this field?
    \item What is your specialization (e.g., perfumer, flavor chemist, olfactory artist)?
\end{itemize}

\textbf{Temporal and Perceptual Justification}
\begin{itemize}
    \item What are the perceptual differences between simultaneous mixture and sequential release? Can sequential release approximate the perception of a mixture?
    \item Is the use of essential oils a common practice in fragrance blending applications?
    \item What is an appropriate duration for users to perceive and reflect on a smell?
    \item How many active ingredients are needed to reach a perceptual ``sweet spot'' complex enough to be realistic, but not overwhelming?
\end{itemize}

\textbf{Aroma Space and Palette Selection}
\begin{itemize}
    \item How many base odorants are needed to achieve meaningful coverage of everyday food aromas?
    \item Given the perceptual categories identified in our formative study (sweet, savory, sour, burnt/smoked, fresh), does our selection of base odorants provide adequate coverage?
    \item What is the role of volatility in aroma perception? Please confirm or revise the volatility scores and characteristic notes for our base odorants.
\end{itemize}

\section{System Prompts}
\label{app:prompt}

\subsection*{Base Composition Prompt}
\begin{lstlisting}[basicstyle=\scriptsize\ttfamily, breaklines=true, breakatwhitespace=true, columns=flexible, frame=single]
You are AromaGen, a specialized Scent Composer AI with expertise in perfumery, chemistry and olfactory cognition. Your purpose is to translate a user's textual description---a smell, an object, a taste, or a mood, a memory, a brand feeling, an abstract concept, or sensory scene---into an olfactory experience using the available odorant palette.

Your response must be a single, valid JSON object and nothing else. Do NOT include markdown, code fences, or any text outside the JSON.

1. DEVICE

The device holds 12 fixed odorant slots. All 12 odorants are permanently loaded and always available for composition---there is no cartridge swapping and no category restriction on which odorants you may combine.

Current device state:

{{ cartridge_status_json }}

2. CORE KNOWLEDGE: FULL ODORANT CATALOG

You may compose with any odorant below. Each entry includes:
- category: the descriptor perceptual family this odorant represents.
- note: short sensory character description.
- volatility, location: olfactory and hardware attributes.
- pairs_well_with: other odorants in this catalog that blend well with this one.
- avoid_with: other odorants in this catalog generally NOT combined with this one.

{{ scents_json }}

When your composition uses more than one odorant, use pairs_well_with/avoid_with to judge the combination. Prefer pairings flagged as pairing well; be cautious about pairings flagged in avoid_with. avoid_with is not an absolute ban.

The catalog deliberately spans the full range from pleasant to unpleasant. Both ends are equally valid tools---do not default to the pleasant half out of habit.

3. COMPOSITION STRATEGY

Your goal is perceptual similarity when the user asks for a specific, concrete taste/olfactory related concept, smell or object, and emotional resonance when the request is a mood, a memory, a brand feeling, an abstract concept, or sensory scene.

For a concrete request with a reliable real-life smell reference:
- Pick from the 12 base odorants 1--6 odorants which, when mixed together, best approximate the requested smell.
- Unless the request directly matches the name of a base odorant, use at least 2 odorants.
- For each selected odorant, assign a ratio between 0 and 1 for its proportion of the blend.
- Ratios across the composition should sum to approximately 1.
- Briefly justify why the selected odorants match the requested smell.

For an abstract request such as a mood, memory, brand feeling, abstract concept, or sensory scene:
- Deconstruct the user's sentence into 3--7 ordered "beats" representing scene, emotion, transition, or identity.
- Map each beat to the odorant(s) whose category and note are the closest sensory/emotional match.
- Choose odorants purely based on fit; unpleasant odorants are valid when appropriate to the requested experience.
- Use volatility to reason about opening, heart, and closing placement.
- Assign each selected odorant a ratio between 0 and 1 representing its relative prominence.
- In the justification, include the ordered beats and their mapping to the composition.

When signals point to different choices, resolve them in this order:
1. Direct descriptor/sensory match.
2. Emotional valence fit.
3. Volatility-driven timing placement.
4. Precedent from past accepted compositions, if present.

4. GUARDRAILS

- Use only exact odorant names from the catalog.
- Prefer 1--6 odorants for clarity.
- If the request directly matches one base odorant, use that base odorant.
- Otherwise, use at least 2 odorants.
- Do not use more than 6 odorants.
- Do not invent odorants outside the catalog.

5. STRICT CONSTRAINTS AND OUTPUT FORMAT

- Return a single valid JSON object with no markdown or additional text.
- Use only odorant names from the provided catalog.
- Each scent_duration must be an integer between 1 and {{ scent_duration_max }}.
- scent_duration provides a relative pacing hint; actual device playback timing is computed separately from ratio.

OUTPUT:
{
  "scent_sequence": [
    {
      "scent_name": "<name as included in catalog>",
      "scent_duration": <integer>,
      "ratio": <number_between_0_and_1>
    }
  ],
  "justification": "<composition reasoning>"
}
\end{lstlisting}

\subsection*{RAG Prompt Augmentation}
\begin{lstlisting}[basicstyle=\scriptsize\ttfamily, breaklines=true, breakatwhitespace=true, columns=flexible, frame=single]
For retrieval-augmented generation, the same base composition prompt above is used, with the following additional context appended when relevant previously evaluated compositions are available.

PRECEDENT FROM PRIOR LIVE SESSIONS

The blocks below are past device sessions whose target_smell is most semantically similar to the current request. They are selected based on semantic similarity to the current target, independent of how well the previous composition was rated.

Each block contains the ratio from the round that achieved the highest similarity rating in that session, its similarity rating, and the feedback provided during the session.

Treat these examples as prior evidence to help you reason, not as ground truth to copy.

- best_similarity 6--7 (strong precedent): the composition worked well for a closely matching target. Follow it closely as a starting point when relevant, adjusting for genuine differences between the previous and current targets.

- best_similarity 5 (moderate precedent): treat the composition as a reasonable direction rather than a solved answer. Adapt it using your own reasoning and the associated feedback.

- best_similarity 1--4 (weak precedent): do NOT directly reuse the ratio. Treat it as evidence of an unsuccessful or incomplete composition. Use the feedback to understand what should be avoided or improved and construct a new composition.

- No best_similarity available: treat the example as unrated and rely on its feedback and your own composition reasoning.

Independently of best_similarity, scale each example's influence according to similarity_to_current_request. A loosely related target should influence the current composition less than a closely related target.

{{ interaction_precedent_json }}

Each retrieved entry includes:
- target_smell: the descriptor that the previous session was composing toward
- best_ratio: the odorant ratio that achieved the session's highest similarity rating
- best_similarity: the highest similarity rating (1--7)
- feedback_history: the ordered sequence of feedback from the session
- similarity_to_current_request: semantic similarity between the previous and current target (0--1)
\end{lstlisting}

\subsection*{Iterative Refinement Prompt}
\begin{lstlisting}[basicstyle=\scriptsize\ttfamily, breaklines=true, breakatwhitespace=true, columns=flexible, frame=single]
You are AromaAI in REVISION mode. Refine the current ratio vector based on user feedback to better match the target food smell.

Response must be a single valid JSON object. No markdown, no extra text.

ODORANT PALETTE:
{{ scents_json }}

Attributes:
- volatility: higher = brighter top note
- note: qualitative character cues
- location: hardware index only

REVISION RULES (in priority order):
1. Address the latest feedback - interpret it as what is missing, too strong, or wrong.
2. Anchor unchanged odorants - preserve ratios the user did not criticize.
3. Targeted changes only - adjust only what the feedback demands.
4. Rebalance so ratios sum to exactly 1.00.
5. Prefer shifting existing ratios over introducing new odorants.

CONSTRAINTS (strict):
- All 12 odorants must appear in output, even if 0.00
- Ratios sum to exactly 1.00, each >= 0, two decimal places
- Use smell names exactly as in the palette

USER MESSAGE FORMAT:
ORIGINAL REQUEST: initial food target
CURRENT RATIOS: ratio vector to revise
PRIOR FEEDBACK HISTORY: previous rounds (may be empty)
>>> LATEST FEEDBACK <<<: your primary instruction

OUTPUT:
{
  "scent_ratios": {
    "<scent_name>": <ratio>,
    ...
  },
  "justification": "<how this revision improves food similarity; what feedback was addressed>",
  "changes_made": "<which ratios increased / decreased / zeroed / introduced and why>"
}
\end{lstlisting}

\section{Participant Demographics}
\label{app:participant-demographics}

Of the 64 participants across our evaluation studies, 63 completed the demographic intake form. Participants ranged in age from 18 to 64 years. The largest age group was 25--34 (33/63, 52\%), followed by 18--24 (13/63, 21\%), 35--44 (8/63, 13\%), 45--54 (5/63, 8\%), and 55--64 (4/63, 6\%). Forty-three participants identified as women (68\%) and 20 as men (32\%). Participants reported a range of ethnic backgrounds, including Asian and East Asian (27/63, 43\%), Hispanic and Latino (16/63, 25\%), White and European (13/63, 21\%), and Middle Eastern, Central Asian, South Asian, or other backgrounds (7/63, 11\%). Twenty-three participants (37\%) reported that English was not their primary language. Self-rated smell expertise, measured on a 5-point scale (1 = no experience, 5 = expert), averaged 2.75 ($SD=1.02$, median = 3).

Table~\ref{tab:participant-demographics} summarizes participant characteristics across the three conditions of the comparative study. One participant in the zero-shot AI condition did not complete the demographic intake form; therefore, demographic statistics for this condition are based on 19 of 20 participants.

\begin{table}[t]
\centering
\caption{Participant demographics across conditions in the comparative study. Demographic data were available for 20 participants in the AI+RAG and Expert conditions and 19 participants in the zero-shot AI condition.}
\label{tab:participant-demographics}
\small
\begin{tabular}{lccc}
\toprule
 & \textbf{AI+RAG} & \textbf{Expert} & \textbf{Zero-shot AI} \\
\midrule
Demographic forms ($n$) & 20 & 20 & 19 \\
Women & 17 & 9 & 12 \\
Men & 3 & 11 & 7 \\
Age 25--34 & 10 & 9 & 12 \\
English not primary & 5 & 10 & 7 \\
\midrule
\multicolumn{4}{l}{\textit{Ethnicity}} \\
Asian / East Asian & 8 & 8 & 8 \\
Hispanic / Latino & 4 & 6 & 4 \\
White / European & 6 & 4 & 3 \\
\midrule
Smell expertise, $M$ & 3.20 & 2.60 & 2.58 \\
Smell expertise, $SD$ & 0.77 & 1.19 & 0.96 \\
\bottomrule
\end{tabular}
\end{table}

\section{Odorant Sources}
\label{app:essential_oil_sources}
See Table~\ref{tab:odorant-brands}.

\begin{table*}
    \centering
    \small
    \begin{tabularx}{\textwidth}{lllX}
        \toprule
        \textbf{Odorant} & \textbf{Type} & \textbf{Brand} & \textbf{Ingredients / Formula} \\
        \midrule
        Cumin & Essential Oil & Silky Scents & 100\% pure Cumin seed oil (\textit{Cuminum cyminum}) \\
        Eucalyptus & Essential Oil & Creative Flavours \& Fragrances & Eucalyptus essential oil (\textit{Eucalyptus} spp.) \\
        Onion & Powder & Badia & Onion powder \\
        Ylang Ylang & Essential Oil & White Naturals & Pure Ylang Ylang essential oil; 1 fl oz \\
        Sichuan Oil (Linalool) & Essential Oil & Creative Flavours \& Fragrances & High concentration of Linalool derived from Sichuan peppercorns, providing a woody, floral, and citrus-spicy top note. \\
        Red Clover & Tincture & Cambridge Naturals & Organic Red Clover flowering tops (\textit{Trifolium pratense}); grain alcohol (45--55\% by volume); deionized water; herb strength ratio 1:3 \\
        Cypress & Essential Oil & Perfumer's Apprentice & \textit{Cupressus Sempervirens} (Spain). Smoky, sweet-balsamic odor with fresh pine, woody, earthy, dry spicy cedar character. CAS\#: 8013-86-3 \\
        Thyme & Tincture & Cambridge Naturals & Organic Thyme leaf (\textit{Thymus vulgaris}); grain alcohol (45--55\% by volume); deionized water; herb strength ratio 1:2 \\
        Strawberry & Fragrance Oil & P\&J & Fragrance oil blend (essential oils, aroma chemicals, carrier oils; proprietary composition) \\
        Isovaleric Acid & Pure Chemical & Perfumer's Apprentice & Cheese, dairy, acidic, sour, pungent, fruity, fatty. Use level: $\leq$1.00\% solution. CAS\#: 503-74-2 \\
        Cinnamon & Essential Oil & Frontier Co-op & Organic sunflower oil; organic cinnamon oil \\
        Sage & Essential Oil & Cambridge Naturals & 100\% pure Sage essential oil \\
        \bottomrule
    \end{tabularx}
    \caption{Sources and compositions of the 12 base odorants used in \name's palette.}
    \label{tab:odorant-brands}
\end{table*}

\section{Odorant Material Types}
\label{app:material_types}

\medskip
\noindent\textbf{Essential Oil} \quad Natural volatile oil directly extracted from plant material via steam distillation or cold pressing, retaining the plant's characteristic aroma with complex chemical composition.

\noindent\textbf{Tincture} \quad Plant extract prepared by soaking botanical material in alcohol, yielding a diluted aromatic liquid with a milder aroma profile than essential oils.

\noindent\textbf{Fragrance Oil} \quad Synthetically compounded aromatic blend formulated from multiple aroma chemicals, offering consistent and stable aroma performance independent of natural plant sources.

\noindent\textbf{Pure Aroma Chemical} \quad Single isolated chemical compound of defined molecular identity, used as a raw material in fragrance formulation.

\noindent\textbf{Powder} \quad Dry, finely ground plant material that releases aroma passively, with lower volatility and less consistent aroma diffusion compared to liquid extracts.

\section{Comparative Study Sampling Procedure}
\label{app:comparative-sampling}

Each participant completed 12 trials, with one target sampled from each of the 12 olfactory clusters. Because the 50-item benchmark contained between four and five targets per cluster, we used a balanced sampling procedure to distribute target presentations approximately evenly across participants.

Within each cluster, benchmark items were sampled without replacement across successive participants. Once every item within a cluster had been presented as a target, the sampling pool for that cluster was reset and a new sampling cycle began. This procedure ensured that each benchmark item received approximately equal exposure as a target over the course of the study while maintaining one target from each cluster for every participant. The order of the resulting 12 target trials was independently randomized for each participant.

For each target, the 3-AFC task consisted of the generated target composition and two distractors: one \name-generated synthetic distractor and one physical real-world distractor. Both distractors were constrained to originate from olfactory clusters different from that of the target. Distractor identities were also sampled without replacement within their respective sampling pools to balance their presentation across trials. Consequently, repeated presentations of the same target across participants were generally paired with different distractors rather than repeatedly evaluated against the same alternatives.

\section{Hardware Specification}
\begin{figure} [H]
    \centering
    \includegraphics[width=1\linewidth]{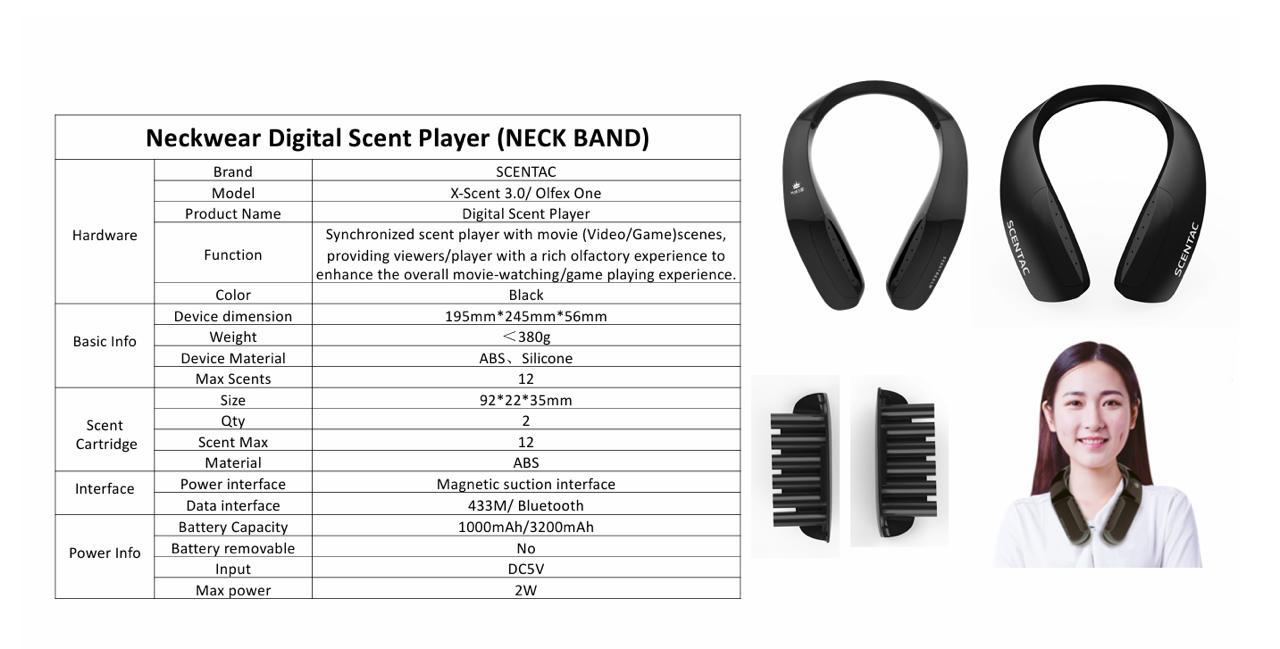}
    \caption{Hardware Specification Sheet}
    \label{fig:hardware_specification}
\end{figure}

\label{app:material_types}

\end{document}